\documentclass[prb,aps,twocolumn]{revtex4}
\usepackage{bm}
\usepackage{amsmath,amssymb}
\usepackage{color,multirow}
\usepackage{graphicx}
\begin{document}
\title{Odd-frequency pair in topological superconductivity of 1D magnetic chain }
\author{Hiromi Ebisu$^1$, Keiji Yada$^2$, Hideaki Kasai$^{1}$, and Yukio Tanaka$^2$}
\affiliation{$^1$Department of Applied Physics, Osaka University, Osaka 565-0871, Japan\\
$^2$Department of Applied Physics, Nagoya University, Nagoya 464-8603, Japan}
\begin{abstract}
A chain of magnetic atoms with non-collinear spin configuration on a superconductor is a promising new system that can host Majorana Fermions (MFs).
In this study, we clarify that in the presence of MFs, an odd-frequency Cooper pair is generated at the edge of the chain.
Furthermore, it is revealed that this feature is robust against the distance between magnetic atoms as far as this distance is shorter than coherence length of the superconductor.
We also elucidate the close relationship between the pair amplitude of the odd-frequency pair and the direction of the MF spin.
If Rashba-type spin-orbit coupling is included, MFs can be realized even in a collinear alignment of magnetic atoms, $i.e.$, in a ferromagnetic or anti-ferromagnetic chain on a superconductor.
Then, the odd-frequency pairing is generated at the edge, similar to the non-collinear case.
Based on our results, it can be concluded that the detection of the zero energy peak of the local density of states by scanning tunneling microscopy at the edge of the magnetic chain is strong evidence for the generation of odd-frequency pairing.
\end{abstract}
\pacs{pacs}
\maketitle
\thispagestyle{empty}
\section{Introduction}\label{sec1}
The concept of topology has become a central issue in condensed matter physics since the proposal of the $\mathbb{Z}_2$ topological insulator.\cite{hasan10}
Owing to the topologically nontrivial wave function, novel properties emerge on the edge or surface of the material.
In addition to the topological insulator, exploration of topological superconductivity has attracted attention.\cite{qi11,alicea12,tanaka12,Fu10,sasaki11}
In topological superconductors, the excitation of the Majorana Fermion (MF), the fermion whose creation and annihilation operators are identical, is imposed on their boundary.\cite{Kitaev01,Fu08}
Since MFs obey non-Abelian statistics, topological superconductivity is important for future application to quantum computing.\cite{Nayak}
It was suggested that $p$-wave spinless superconductors can show topological superconductivity.\cite{Kitaev01}
However, this type of superconductor has not been discovered thus far.
In this respect, the major challenge in many theoretical and experimental studies is to realize topological superconductivity with MFs in an effective spinless $p$-wave system.\cite{Fu08,Fujimoto09,STF10,Jay,alicea10,lutchyn10,oreg10,lucignano1,lucignano2}
Further, charge transport in these systems was intensively studied.\cite{Fu09,Akhmerov09,Tanaka09,law09,Linder10}

One possible system that can realize topological superconductivity is a semiconducting nanowire deposited on a spin-singlet $s$-wave superconductor with a Zeeman field.\cite{alicea10,lutchyn10}
By tuning the Zeeman field, one can produce a ``spinless'' system in which the degree of spin is halved on the Fermi surface.
Many previous studies were aimed at verifying the topological superconductivity through charge transport experiments to probe the peak of the zero-energy spectrum that stems from the existence of MFs.\cite{Deng,Rokhinson,Das,Mourik,Eduardo}
However, the origin of the zero-energy spectrum has been attributed to several other factors as well.\cite{Liu2012,Dasnano,bagrets,pikulin,kells,hlee}
In addition, energy-gap closing due to the topological phase transition has not been observed yet.

Another study proposed that a chain of magnetic atoms on the top of the $s$-wave superconductor can host MFs.\cite{Choy2011}
A magnetic atom forms the so-called Shiba state within the induced gap in the superconductor system,\cite{Shiba,Yu} and the 1D array of these moments on the top of the $s$-wave superconductor can be ``spinless'' which leads to a topologically non-trivial state.\cite{Yazdani2013}
The advantage of this magnetic chain is that one can access each magnetic atom by scanning tunneling microscopy (STM); thus, MFs can be spatially probed.
There have been several theoretical investigations on this magnetic chain,\cite{Martin2012,Klinovaja2013,Braunecker,Nakosai2013,Kim2014,Pientka2014,Ojanen2014,Ojanen2,Oppen,Brydon,heimes} and experimental confirmation of topological superconductivity in this system is highly promising.\cite{yazdaniex}

It has been found that odd-frequency pairing\cite{Berezinskii,Balatsky92,Coleman} has a close connection with topological-superconductivity-hosting surface Andreev bound state (ABS).\cite{tanaka12,odd1,odd3,odd3b,wakatsuki}
The concept of the odd-frequency pairing naturally arises on introducing the frequency dependence of the pairing function.
The odd-frequency pair ubiquitously exists in inhomogeneous superconducting systems with broken symmetries in, for instance, spin rotation\cite{Efetov1,Efetov2,fominov1,eschrig_natuer,yokoyama1,halterman1,Asano1} or translation.\cite{odd3,odd3b,Eschrig2007,Yokoyamavortex,Mizushima,Bakurskiy,Balatsky1,Balatsky2,Higashitani,Asano2014}
In these systems, the odd-frequency pairings are prominent in the presence of the surface ABS.\cite{ABS,ABSb,Hu,TK95,kashiwaya00}
Since the MF is a type of ABS, it is expected to be related to odd-frequency pairing.
It is elucidated that the odd-frequency pair becomes prominent at the edge of the nanowire/$s$-wave superconductor junction described above.\cite{Asano2013,Stanev}
The system consisting of a magnetic chain on an $s$-wave superconductor is useful for developing an understanding of the relation between the MF and odd-frequency pairing.

In this study, we reveal that the odd-frequency pairing becomes prominent at the edge of the array of non-collinear spins of magnetic atoms corresponding to MFs.
We find a close relationship between the pair amplitude of the odd-frequency pairing and the direction of the MF's spin.
These conclusions are independent of the interval between magnetic atoms as far as this interval is shorter than coherence length of the superconductor.
If Rashba-type spin-orbit coupling (SOC) is included, ferromagnetic and anti-ferromagnetic configurations can also be topologically non-trivial, and odd-frequency pairing is generated near the edge.
This paper is organized as follows.
In Sec. \ref{sec2}, we introduce various symmetry of the pair potential and clarify a general relationship between odd-frequency pair amplitude and MF. 
In Sec. \ref{sec3}, we focus on a Shiba state with a single magnetic atom on an $s$-wave superconductor.
In Sec. \ref{sec4}, we investigate an array of magnetic atoms on an $s$-wave superconductor (ferromagnetic, anti-ferromagnetic, and non-collinear configurations).
The relation between the amplitude of the odd-frequency pair, parity of the wave function of the Bogoliubov de Gennes Hamiltonian, and direction of the MF spin is discussed in detail.
In Sec.  \ref{sec3} and \ref{sec4}, we do not assume SOC, whereas in Sec. \ref{sec5}, we include the SOC and find that ferromagnetic and anti-ferromagnetic chains can host MFs and odd-frequency pairing.

\section{General argument on symmetries of pair potential}\label{sec2}
In this paper, we investigate the local density of states (LDOS) and pair amplitudes
of the 1D array of magnetic atoms on a superconductor.
Before going to the detailed calculations, we review the various types of 
pair potential by introducing Matsubara Green's function. The relationship between odd-frequency pair and MF is also provided.\par
We consider a tight-binding BdG Hamiltonian with site 
$j$ and spin index $\sigma$. 
More detailed discussion will be given below 
Eqs.(18) and (19). 

Retarded Green's function and Matsubara Green's function can be defined as follows:
\begin{equation}
G^{\text{R}}(E,j\sigma,j^{\prime}\sigma^{\prime})=\Bigl(\frac{1}{E+i\epsilon-\mathcal{H}}\Bigr)_{j\sigma,j^{\prime}\sigma^{\prime}}=\left( \begin{array}{cc}
G^{\text{R}} & F^{\text{R}} \\
\tilde{F}^{\text{R}} & \tilde{G}^{\text{R}} 
\end{array} \right)\label{green}
\end{equation}
\begin{equation}
G(\omega_n,j\sigma,j^{\prime}\sigma^{\prime})=\Bigl(\frac{1}{i\omega_n-\mathcal{H}}\Bigr)_{j\sigma,j^{\prime}\sigma^{\prime}}=\left( \begin{array}{cc}
G & F \\
\tilde{F} & \tilde{G} 
\end{array} \right),
\end{equation}
where $\epsilon$ and $\omega_n$ are infinitesimal positive number and Matsubara frequency, respectively.
Throughout this paper, we fix $\omega_n$ and $\epsilon$ as $\omega_n/t=0.01$, $\epsilon/t=0.001$.
The pair amplitude of the Cooper pair is described by the anomalous part of the Matsubara Green's function.
This pair amplitude is classified into four kinds of symmetry with respect to frequency, spin and parity:
even-frequency spin-singlet even-parity (ESE),
even-frequency spin-triplet odd-parity (ETO),
odd-frequency spin-triplet even-parity (OTE),
and odd-frequency spin-singlet odd-parity (OSO).
Since the pair potential in the present model is BCS $s$-wave pairing,
the most dominant pairing is $s$-wave ($p$-wave) one for even-parity (odd-parity) pairing.
Thus, we focus on these pairings.
LDOS and corresponding pair amplitudes at the position $j$ are given by
\begin{equation}
\rho(E,j)=-\frac{1}{\pi}\sum_{\sigma}\text{Im}G^{\text{R}}(E,j\sigma,j\sigma),
\end{equation}
\begin{equation}
f^{\text{s}\;\text{odd(even)}}_{\sigma,\sigma^{\prime}}=\frac{\tilde{F}(\omega_n,j\sigma,j\sigma^{\prime})-(+)\tilde{F}(-\omega_n,j\sigma,j\sigma^{\prime})}{2},
\end{equation}
\begin{eqnarray}
f^{\text{p}\;\text{odd(even)}}_{\sigma,\sigma^{\prime}}=\frac{1}{2}\Bigl[\frac{\tilde{F}(\omega_n,j+1\sigma,j\sigma^{\prime})-\tilde{F}(\omega_n,j\sigma,j+1\sigma^{\prime})}{2}\notag\\
-(+)\frac{\tilde{F}(-\omega_n,j+1\sigma,j\sigma^{\prime})-\tilde{F}(-\omega_n,j\sigma,j+1\sigma^{\prime})}{2}\Bigr].\label{odd}
\end{eqnarray}
The anomalous part of the Matsubara Green's function is also written as
\begin{equation}
\tilde{F}(\omega_n,j\sigma,j^{\prime}\sigma^{\prime})=\sum_{\lambda}\Bigl\{\frac{v_{\lambda}(j\sigma)u^{*}_{\lambda}(j^{\prime}\sigma^{\prime})}{i\omega_n-E_{\lambda}}+\frac{u^{*}_{\lambda}(j\sigma)v_{\lambda}(j^{\prime}\sigma^{\prime})}{i\omega_n+E_{\lambda}}\Bigr\}
\end{equation}
where the eigenenergy of the BdG Hamiltonian is denoted by $E_{\lambda}$ and 
$u_{\lambda}(j\sigma)(v_{\lambda}(j\sigma))$ is a component of the eigenvectors at the position $j$ in electron (hole) space with spin $\sigma$.
In this representation, $s$-wave odd-frequency pair amplitude at $j$th site becomes 
\begin{eqnarray}
f^{\text{s}\;\text{odd}}_{\sigma\sigma^{\prime}}=\frac{1}{2}[\tilde{F}(\omega_n,j\sigma,j\sigma^{\prime})-\tilde{F}(-\omega_n,j\sigma,j\sigma^{\prime})]\notag\\
=\sum_{\lambda}\frac{-i\omega_n}{\omega_n^2+E_{\lambda}^2}\Bigl(u^{*}_{\lambda}(j\sigma)v_{\lambda}(j\sigma^{\prime})+u^{*}_{\lambda}(j\sigma^{\prime})v_{\lambda}(j\sigma)\Bigr)\label{def}.
\end{eqnarray}
In the following discussion, we focus on the case when the system is topologically non-trivial with zero energy edge state. 
The most dominant term  in Eq. (\ref{def}) comes from 
zero energy state with $E_0=0$ since the denominator becomes minimum. 
Thus, we approximate $f^{\text{s}\;\text{odd}}_{\sigma\sigma^{\prime}}$ as
\begin{equation} 
 f^{\text{s}\;\text{odd}}_{\sigma\sigma^{\prime}}\approx\frac{1}{i\omega_n}\Bigl(u^{*}_{0}(j\sigma)v_{0}(j\sigma^{\prime})+u^{*}_{0}(j\sigma^{\prime})v_{0}(j\sigma)\Bigr).\label{apr}
\end{equation} 
In other words, odd-frequency pair amplitude has a close relationship with the wave function of the zero energy states. 
Now we focus on the right edge $j=L_{r}$. When the system is topologically non-trivial, zero energy state can be described as
\begin{equation}
\Psi=(u_0(L_r\uparrow),u_0(L_r\downarrow),v_0(L_r\uparrow),v_0(L_r\downarrow))^T.
\end{equation}
Majorana operator is given as
\begin{eqnarray}
\gamma^{\dagger}=u_0(L_r\uparrow)c_{L_r\uparrow}^{\dagger}+u_0(L_r\downarrow)c_{L_r\downarrow}^{\dagger}\notag\\
+v_0(L_r\uparrow)c_{L_r\uparrow}+u_0(L_r\downarrow)c_{L_r\downarrow}
\end{eqnarray}
where $c_{L_r\sigma}^{\dagger}(c_{L_r\sigma})$ represents creation 
(annihilation) operator at $j=L_r$. To satisfy $\gamma^{\dagger}=\gamma$, we have following condition\cite{tanaka12}:
\begin{eqnarray}
u_0(L_r\uparrow)=v_0^{*}(L_r\uparrow)\notag\\
u_0(L_r\downarrow)=v_0^{*}(L_r\downarrow).\label{special} 
\end{eqnarray}
Using this condition, we relate the pair amplitudes of 
odd-frequency $s$-wave pair to the spin of MF. 
At $j=L_r$, 
following relations are satisfied, 
\begin{equation} 
f^{\text{s}\;\text{odd}}_{\sigma\sigma^{\prime}}\approx\frac{1}{i\omega_n}\Bigl(u^{*}_{0}(L_r\sigma)v_{0}(L_r\sigma^{\prime})+u^{*}_{0}(L_r\sigma^{\prime})v_{0}(L_r\sigma)\Bigr), 
\end{equation}
\begin{equation}
\frac{f^{\text{s}\;\text{odd}}_{\uparrow\uparrow}-f^{\text{s}\;\text{odd}}_{\downarrow\downarrow}}{f^{\text{s}\;\text{odd}}_{\uparrow\downarrow}}=\frac{2(|u_0(L_r\uparrow)|^2-|u_0(L_r\downarrow)|^2)}{u_0^{*}(L_r\uparrow)u_0(L_r\downarrow)+u_0^{*}(L_r\downarrow)u_0(L_r\uparrow)}
\end{equation}
On the other hand, the expectation value of the spin at zero energy, $i.e.$ the spin of MF\cite{sticlet,He1,Y.Nagai} is given by
\begin{eqnarray}
\left\langle s_x \right\rangle&=&\frac{1}{2}\left(u_0^{*}(L_r\uparrow),u_0^{*}(L_r\downarrow)\right)\left(
\begin{array}{cc}
0&1\\
1&0
\end{array}
\right)
\left( \begin{array}{c}
u_0(L_r\uparrow)\\
u_0(L_r\downarrow)
\end{array} \right)\notag\\
&=&\frac{1}{2}(u_0^{*}(L_r\uparrow)u_0(L_r\downarrow)+u_0^{*}(L_r\downarrow)u_0(L_r\uparrow))
\end{eqnarray}
\begin{eqnarray}
\left\langle s_y \right\rangle&=&\frac{1}{2}\left(u_0^{*}(L_r\uparrow),u_0^{*}(L_r\downarrow)\right)\left(
\begin{array}{cc}
0&-i\\
i&0
\end{array}
\right)
\left( \begin{array}{c}
u_0(L_r\uparrow)\\
u_0(L_r\downarrow)
\end{array} \right)\notag\\
&=&\frac{i}{2}(-u_0^{*}(L_r\uparrow)u_0(L_r\downarrow)+u_0^{*}(L_r\downarrow)u_0(L_r\uparrow))\label{sy}
\end{eqnarray}
\begin{eqnarray}
\left\langle s_z \right\rangle&=&\frac{1}{2}\left(u_0^{*}(L_r\uparrow),u_0^{*}(L_r\downarrow)\right)\left(
\begin{array}{cc}
1&0\\
0&-1
\end{array}
\right)
\left( \begin{array}{c}
u_0(L_r\uparrow)\\
u_0(L_r\downarrow)
\end{array} \right)\notag\\
&=&\frac{1}{2}((|u_0(L_r\uparrow)|^2-|u_0(L_r\downarrow)|^2))
\end{eqnarray}
Therefore, we find that odd-frequency pair amplitudes and the 
ratio of $\left\langle s_z \right\rangle$ and $\left\langle s_x \right\rangle$ satisfy the following relation,
\begin{equation}
\frac{f^{\text{s}\;\text{odd}}_{\uparrow\uparrow}-f^{\text{s}\;\text{odd}}_{\downarrow\downarrow}}{f^{\text{s}\;\text{odd}}_{\uparrow\downarrow}}=2\frac{\left\langle s_z \right\rangle}{\left\langle s_x \right\rangle}.\label{conc}
\end{equation}
In later sections, using Eq.(\ref{apr}), we discuss how odd-frequency pair relates with parity of the wave function by our numerical results. Further, we confirm Eq.(\ref{conc}) which shows the relation between the 
 odd-frequency pair amplitude and the direction of the spin of MF. 
\section{Single magnetic atom on a superconductor}\label{sec3}
\begin{figure}[h]
\begin{center}
\includegraphics[width=9.5cm]{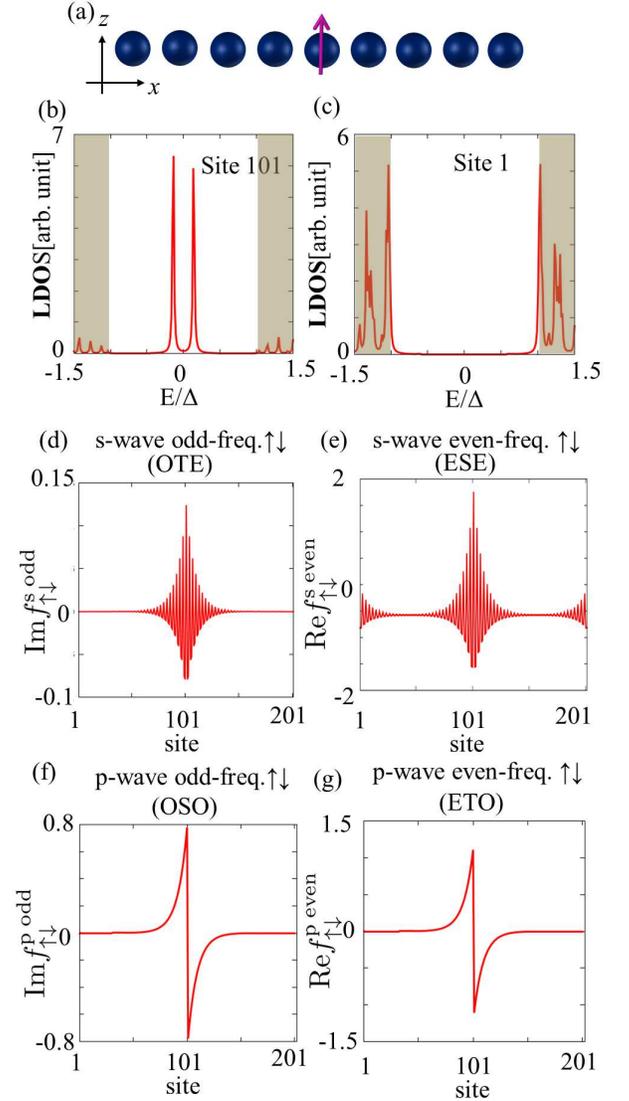}
\caption{
(a)A schematic picture of the single magnetic moment on an $s$-wave superconductor.
(b)(c)LDOS at site 101 and site 1 in the case of the single magnetic moment of $s$-wave superconductor.
The magnetic moment is located at site 101 in the chain with the length of 201. The shaded area represents continuum level in the system with infinite length of the chain. (d), (e), (f), and (g) represent pair amplitudes of 
$s$-wave odd-frequency, $s$-wave even-frequency, $p$-wave odd-frequency, and $p$-wave even-frequency pairing, respectively.
Parameters are chosen as $\mu/t=-1, \Delta/t=0.1,$ and $J/t=2.0$.
}\label{fig1}
\end{center}
\end{figure}
\begin{figure}[h]
\begin{center}
\includegraphics[width=9.5cm]{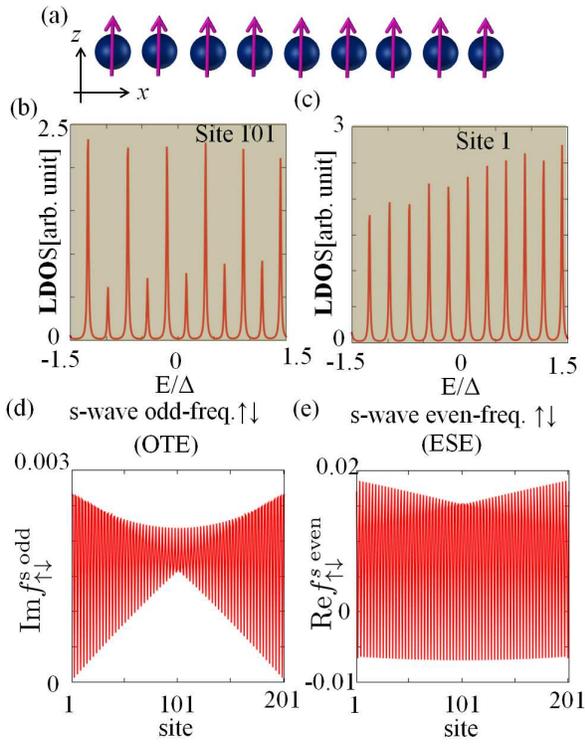}
\caption{
(a)A schematic picture of the ferromagnetic chain on a spin-singlet $s$-wave superconductor.
(b)(c)LDOS of chain at the site 101 and 1. The length of the chain is set to be 201. The shaded area represents continuum level in the system with infinite length of the chain.
(d)(e) pair amplitudes of $s$-wave odd-frequency and even-frequency pair.
Parameters are chosen as $\mu/t=-1, \Delta/t=0.1,$ and $J/t=2.0$.}\label{fig2}
\end{center}
\end{figure}
\begin{figure}[h]
\begin{center}
\includegraphics[width=9.5cm]{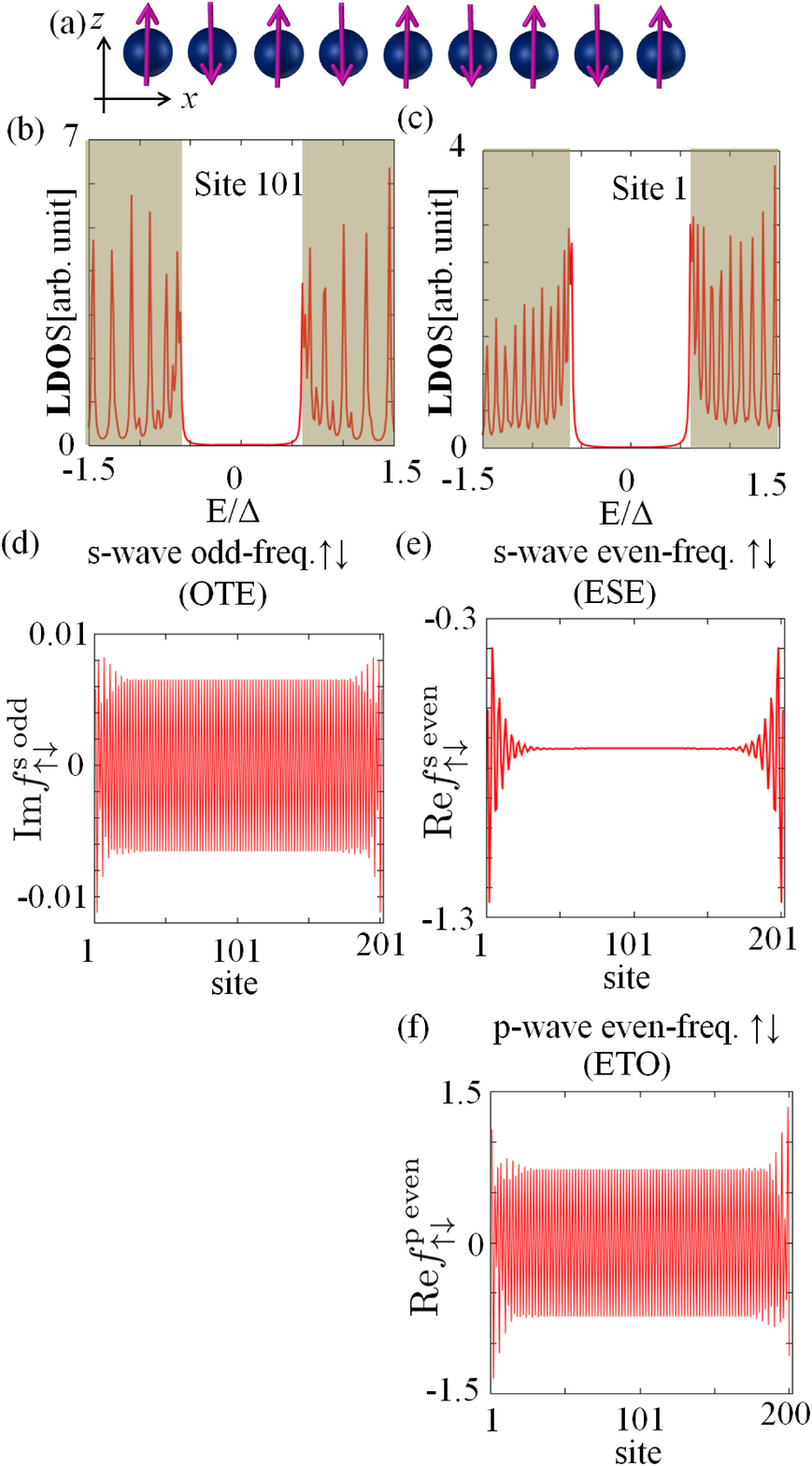}
\caption{
(a)A schematic picture of the anti-ferromagnetic chain on $s$-wave superconductor.
(b)(c)LDOS of this chain at the site 101 and 1. The length of the chain is 201. The shaded area represents continuum level in the system with infinite length of the chain.
(d),(e), and (f) represent pair amplitudes of $s$-wave odd-frequency, $s$-wave even-frequency,
and $p$-wave even-frequency pairing, respectively.
Parameters are chosen as $\mu/t=-2.5, \Delta/t=0.1,$ and $J/t=2.0$.}\label{fig3}
\end{center}
\end{figure}
Now we calculate LDOS and pair amplitude of 1D array of magnetic chain.
This magnetic chain can be regarded as the chain of Shiba state.
Thus, we first look at LDOS and pair amplitudes in the case with
single magnetic atom on $s$-wave superconductor as depicted in Fig.1(a).
The BdG Hamiltonian for this 1D tight-binding model is given by
\begin{eqnarray}
\mathcal{H}&=&-t\sum_{\left\langle i,j \right\rangle\sigma}c^{\dagger}_{i\sigma}c_{j\sigma}-\mu\sum_{i,\sigma}c^{\dagger}_{i\sigma}c_{i\sigma}\notag\\
&+&\sum_i\Bigl( \Delta c^{\dagger}_{i\uparrow}c^{\dagger}_{i\downarrow}+\text{H.c.}\Bigr)\notag\\
&+&J\sum_{\sigma,\sigma^{\prime}}c^{\dagger}_{i_0\sigma}(\sigma_z)_{\sigma,\sigma^{\prime}}c_{i_0\sigma^{\prime}},\label{model} 
\end{eqnarray}
where $c^{\dagger}_{i\sigma}(c_{i\sigma})$ is a creation (an annihilation) operator for an electron with position $i$ and spin $\sigma$. $t$, $\mu$, $\Delta$, and $J$ represent hopping between nearest neighbor sites $\left\langle i,j \right\rangle$, chemical potential, pair potential, and exchange coupling \cite{remark}, respectively.
In this model, magnetic moment is put at site $i_0$.

In Fig.1, we show LDOS and pair amplitudes decomposed into four types of pairings. 
As shown in Figs.1(b) and (c), we can see that there is localized state within the superconducting energy gap in the LDOS at around the magnetic atom,
which is called Shiba state.\cite{Shiba}
At around the magnetic atom, $s$-wave odd-frequency pairing amplitude is produced as shown in Fig.1(d).
In addition, $s$-wave even-frequency pair amplitude has a 
spatial modulation as shown in Fig.1(e).
These $s$-wave even-frequency pair amplitude shows spatially oscillating behavior. This oscillation stems from the Friedel oscillation and   
the order of the period of this oscillation is $1/k_{\text{F}}$ with Fermi wave number $k_{F}$. The periodicity $1/k_{\text{F}}$ is roughly estimated as the length of the one site, which well agrees with our numerical results. 
Other than $s$-wave pairings described above,
$p$-wave odd-frequency and $p$-wave even-frequency pairings 
are generated around the magnetic moment as seen from Figs.1(f) and (g).
It is remarkable that all four types of pair amplitudes exist
in the presence of magnetic atom.
However, the behaviors of these pairings are different each other.
Since we consider the $s$-wave pair potential,
ESE pairing exists overall the chain.
The other three pairings (OTE, OSO, ETO) are induced by the magnetic atom,
and therefore, they locate near the magnetic atom. 
In the present system,
both translational and spin-rotational symmetries are broken.
Due to the translational symmetry breaking, 
OSO pairing is generated around the magnetic atom.\cite{odd3,odd3b,tanaka12}
Furthermore, due to the breakdown of the spin-rotational symmetry, 
OTE and ETO pairings are generated.\cite{tanaka12,yokoyama1}
However, since there is no spin flip scattering process, 
the induced pair has $S_{z}=0$.\cite{yokoyama1}
Thus, all of the pair amplitudes have spin index with 
$\uparrow\downarrow$ or $\downarrow\uparrow$ and 
equal-spin triplet pair amplitudes, 
$e.g.$ $f^{\text{s}\;\text{odd}}_{\uparrow\uparrow}$, are absent. 
\section{Magnetic chain on a superconductor}\label{sec4}
In this section, we focus on the array of the magnetic atoms on the top of an $s$-wave superconductor.
We analyze ferromagnetic and anti-ferromagnetic configurations in subsection A, and non-collinear one in subsection B.
\subsection{Ferromagnetic and anti-ferromagnetic chain}
In this subsection, we consider the magnetic chains
with ferromagnetic (Fig.2(a)) and anti-ferromagnetic configurations (Fig.3(a)). 
Then, the exchange term in Eq.(1) is changed to $J\sum_{i,\sigma,\sigma^{\prime}}c^{\dagger}_{i\sigma}(\sigma_z)_{\sigma,\sigma^{\prime}}c_{i\sigma^{\prime}}$ for the ferromagnetic case and $\sum_{i,\sigma,\sigma^{\prime}}(Jc^{\dagger}_{2i-1\sigma}(\sigma_z)_{\sigma,\sigma^{\prime}}c_{2i-1\sigma^{\prime}}-Jc^{\dagger}_{2i\sigma}(\sigma_z)_{\sigma,\sigma^{\prime}}c_{2i\sigma^{\prime}})$ for the anti-ferromagnetic case. 
In Figs.2(b)$\sim$(e) and Figs.3(b)$\sim$(f), we plot LDOS and pair amplitudes in the case of ferromagnetic and anti-ferromagnetic chain, respectively.
In the ferromagnetic case, bulk energy gap closes for $|J|>|\Delta|$. 
Thus, in the finite chain, oscillating behavior of $s$-wave even-frequency pair amplitude which is originating from the broken translational symmetry at the edge spreads to the entire chain as shown in Fig.2(e).
In the present system,
the induced odd-frequency $p$-wave pair amplitude is negligible
similar to the edge of spin-singlet $s$-wave without magnetic order.\cite{odd3,odd3b}
On the other hand, due to the breakdown of spin-rotational symmetry, 
$s$-wave odd-frequency pair can be realized as shown in Fig.2(d).

Next, we consider the anti-ferromagnetic chain.
Unlike the ferromagnetic chain,
bulk energy spectrum for the anti-ferromagnetic chain has an energy gap.
However, no inner gap states is found in Figs.3 (b) and (c).
As for the pair amplitudes,
$s$-wave odd-frequency and $p$-wave even-frequency pair amplitudes exist the entire chain (see Figs.3 (d) and (f)).
We have confirmed that they become nonzero in the infinite system and their signs are opposite between adjacent sites (See appendix B).
The magnitude of $s$-wave even-frequency pair amplitude is almost constant in the middle chain while it oscillates at the edge (Fig.3 (e)).

In both cases of ferromagnetic and anti-ferromagnetic chain,
although the Hamiltonian has only the ESE pair potential,
there exist the other types of pairings stemming from the symmetry breaking
such as translational or spin-rotational symmetry. 
In Appendix A (B), we analyze the symmetries of the pair amplitudes of the infinite ferromagnetic (anti-ferromagnetic) chain.
These results tell that all the induced pair amplitudes in the finite chain
also exist in the infinite chain.
As we will show in the later sections,
these two configurations cannot be topologically non-trivial
where no inner gap state is localized at the edge.
Therefore, all the induced pair amplitudes are specific to those in the bulk.

\subsection{Non-collinear magnetic chain}
In this subsection, we consider a chain with non-collinear magnetic atoms on $s$-wave superconductor.
Topological properties of this magnetic chain have been discussed in several preexisting papers,\cite{Choy2011,Yazdani2013,Ojanen2014,Brydon}
however, the case when the magnetic atoms are positioned at intervals (the distance between magnetic atoms is more than one sites) has not been considered.
We distinguish these two cases by using the word ``non-separate'' and ``separate''.
We provide topological property of ``separate'' chain in detail.
Further, using equations in Sec.2, the relation between the odd-frequency pair and MF of this chain is clarified in this subsection.\par
Model Hamiltonian of ``non-separate'' chain is 
\begin{eqnarray}
\mathcal{H}&=&-t\sum_{\left\langle i,j \right\rangle\sigma}c^{\dagger}_{i\sigma}c_{j\sigma}-\mu\sum_{i,\sigma}c^{\dagger}_{i\sigma}c_{i\sigma}\notag\\
&+&\sum_i \Bigl(\Delta c^{\dagger}_{i\uparrow}c^{\dagger}_{i\downarrow}+\text{H.c.}\Bigr)+\sum_{\sigma,\sigma^{\prime}}c^{\dagger}_{i\sigma}(J_i)_{\sigma,\sigma^{\prime}}c_{i\sigma^{\prime}}\label{model2}, 
\end{eqnarray}
where $(J_i)_{\sigma,\sigma^{\prime}}$ represents magnetic moment at site $i$.
Here, we consider the non-collinear magnetic moments where the direction of the spin ``rotates'' clockwise in $x$-$z$ plane with a period $N_p$ (see Fig.4(a)).
Then, $(J_i)_{\sigma,\sigma^{\prime}}$ is given by
\begin{equation}
(J_i)_{\sigma,\sigma^{\prime}}=J\Bigl(\cos\theta_i(\sigma_z)_{\sigma\sigma^{\prime}}+\sin\theta_i(\sigma_x)_{\sigma\sigma^{\prime}}\Bigr),
\end{equation}
where $\theta_i=\frac{2\pi}{N_p}n_i $ ($N_p$, $n_i$: integer) denotes the angle from the $z$-axis.
In this paper, we focus on the case $N_p=4$.

Now, we discuss the topological number in the present system
following  Refs. \onlinecite{Kitaev01} and \onlinecite{Yazdani2013}. 
By the gauge transformation,
\begin{equation}
\left( \begin{array}{c}
c_{i\uparrow}\\
c_{i\downarrow}
\end{array} \right)=
U_i\left( \begin{array}{c}
f_{i\uparrow}\\
f_{i\downarrow}
\end{array} \right)=
\left( \begin{array}{cc}
\cos(\theta_i/2)&-\sin(\theta_i/2)\\
\sin(\theta_i/2)&\cos(\theta_i/2)
\end{array} \right)
\left( \begin{array}{c}
f_{i\uparrow}\\
f_{i\downarrow}
\end{array} \right)\label{gauge},
\end{equation}
the corresponding BdG Hamiltonian is written as follows:
\begin{eqnarray}
\mathcal{H}&=&-t\sum_{i\sigma\sigma^{\prime}}f^{\dagger}_{i\sigma}U_i^{\dagger}U_{i+1}f_{i+1\sigma^{\prime}}-t\sum_{i\sigma\sigma^{\prime}}f^{\dagger}_{i+1\sigma}U_{i+1}^{\dagger}U_{i}f_{i\sigma^{\prime}}\notag\\
&-&\mu\sum_{i,\sigma}f^{\dagger}_{i\sigma}f_{i\sigma}+\sum_i \Bigl(\Delta f^{\dagger}_{i\uparrow}f^{\dagger}_{i\downarrow}+\text{H.c.}\Bigr)\notag\\
&+J&\sum_{\sigma,\sigma^{\prime}}f^{\dagger}_{i\sigma}(\sigma_z)_{\sigma,\sigma^{\prime}}f_{i\sigma^{\prime}}.
\end{eqnarray}
$U_i^{\dagger}U_{i+1}$ is calculated as
\begin{equation}
U_i^{\dagger}U_{i+1}=
\left( \begin{array}{cc}
\cos(\theta/2)&-\sin(\theta/2)\\
\sin(\theta/2)&\cos(\theta/2)
\end{array} \right),
\end{equation}
with $\theta=\frac{2\pi}{N_p}$.
Topological number can be obtained by the Pfaffian of the bulk BdG Hamiltonian. We decompose Fermion operators into MF ones
\begin{equation}
f^{(\dagger)}_{j\sigma}=\frac{1}{2}(a_{2j-1\sigma}+(-)ia_{2j\sigma}).\label{MF}
\end{equation}
By performing the Fourier transformation,
\begin{equation}
a_{2j-1(2j)\sigma}=\frac{1}{\sqrt{N_x}}\sum_{k_x}a_{1(2)k_x\sigma}e^{-i(jk_x)},
\end{equation}
we get anti-symmetric Hamiltonian whose basis are $a_{k_x}=$($a_{1k_x\uparrow}$,$a_{1k_x\downarrow}$,$a_{2k_x\uparrow}$,$a_{2k_x\downarrow}$)$^{\text{T}}$,
\begin{equation}
\mathcal{H}=\frac{i}{4}\sum_{k_x}a_{k_x}^{\dagger}\mathcal{H}^{\text{MF}}(k_x)a_{k_x}\label{PF}.
\end{equation}
Nonzero matrix elements of $\mathcal{H}^{\text{MF}}(k_x)$ are 
\begin{eqnarray}
\Bigl(\mathcal{H}^{\text{MF}}(k_x)\Bigr)_{1,3}=-2t\cos(\theta/2)\cos(k_x)-\mu+J,\notag\\
\Bigl(\mathcal{H}^{\text{MF}}(k_x)\Bigr)_{2,4}=-2t\cos(\theta/2)\cos(k_x)-\mu-J,\notag\\
\Bigl(\mathcal{H}^{\text{MF}}(k_x)\Bigr)_{1,4}=-2ti\sin(\theta/2)\sin(k_x)-\Delta,\notag\\
\Bigl(\mathcal{H}^{\text{MF}}(k_x)\Bigr)_{2,3}=2ti\sin(\theta/2)\sin(k_x)+\Delta.
\end{eqnarray}
Then, we obtain the Pfaffian of this matrix as 
\begin{eqnarray}
\text{Pf}[\mathcal{H}^{\text{MF}}(k_x)]&=&-\Bigl(\mathcal{H}^{\text{MF}}(k_x)\Bigr)_{1,3}\cdot\Bigl(\mathcal{H}^{\text{MF}}(k_x)\Bigr)_{2,4}\notag\\
&+&\Bigl(\mathcal{H}^{\text{MF}}(k_x)\Bigr)_{1,4}\cdot\Bigl(\mathcal{H}^{\text{MF}}(k_x)\Bigr)_{2,3}\notag\\
&=&\Bigl(2t\cos(\theta/2)\cos(k_x)+\mu\Bigr)^2-J^2\notag\\
&+&\Bigl(2ti\sin(\theta/2)\sin(k_x)+\Delta\Bigr)^2.
\end{eqnarray}
$\mathbb{Z}_2$ topological number $\nu$ is given by the sign of the product of the Pfaffian at $k_x=0$ and $k_x=\pi$ 
\begin{equation}
(-1)^{\nu}=\text{sgn}\Bigl(\text{Pf}[\mathcal{H}^{\text{MF}}(k_x=0)] \Bigr)\cdot\text{sgn}\Bigl(\text{Pf}[\mathcal{H}^{\text{MF}}(k_x=\pi)] \Bigr).
\end{equation} 
Finally, the parameter region where $(-1)^{\nu}$ is topologically non-trivial
is found to be,
\begin{eqnarray}
\sqrt{(|2t\cos(\theta/2)|-|\mu|)^2+\Delta^2}<|J|\notag\\
<\sqrt{(|2t\cos(\theta/2)|+|\mu|)^2+\Delta^2}. \label{nontrivial}
\end{eqnarray}
Topological phase diagram of non-collinear and ``non-separate'' magnetic chain
is shown in Fig.4(c).
Note that this condition is valid only when the bulk energy gap is non-zero. 
Equation (\ref{nontrivial}) gives us the reason why the ferromagnetic chain ($\theta=0$) and anti-ferromagnetic chain ($\theta=\pi$) cannot be topologically non-trivial.
In the case of the ferromagnetic chain,
the bulk energy gap is closed for $|J|>\Delta$.
Thus, the topological number is well-defined only for $|J|<\Delta$.
In this case, the condition (\ref{nontrivial}) is never satisfied.
In the case of the anti-ferromagnetic chain,
the left hand side and the right hand side in Eq.(\ref{nontrivial})
are identical.
Thus, the anti-ferromagnetic chain cannot be non-trivial.

\begin{figure}[h]
\begin{center}
\includegraphics[width=10cm]{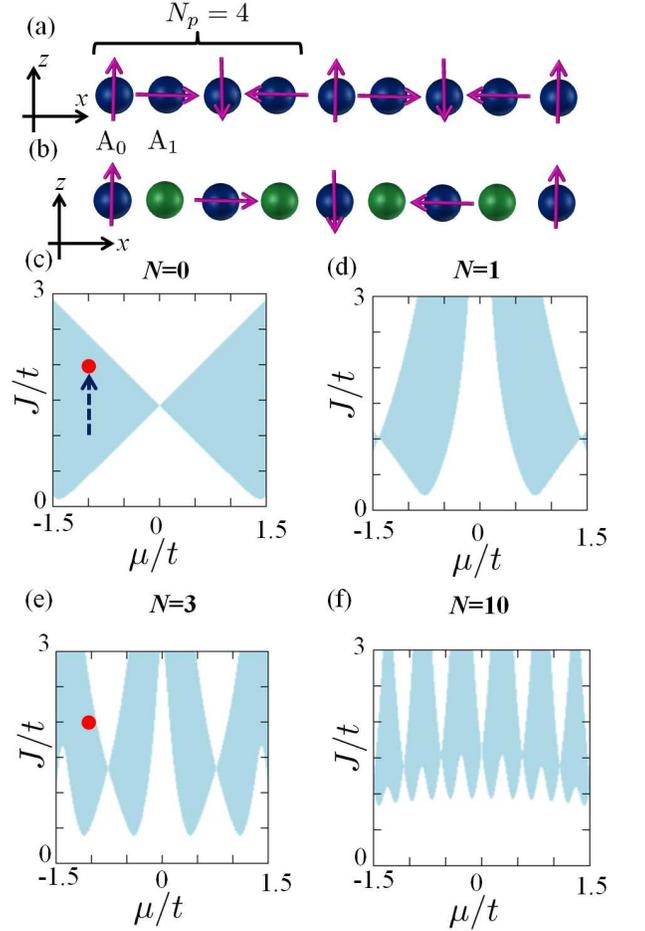}
\caption{
(a)(b)A schematic picture of the non-collinear magnetic chain on $s$-wave superconductor for ``non-separate'' case(a) and for ``separate'' case (b) with vacant sites of $N=1$.
Tilting angle of the magnetic atoms is $\theta=\pi/2$, $i.e.$ $N_p=4$.
(c)$\sim$(f) Phase diagram of the non-collinear magnetic chain in the case of $\Delta/t=0.1$ and $N_p=4$.
The shaded region represents topologically non-trivial one.
}\label{fig4}
\end{center}
\end{figure}
We next focus on the case of the ``separate'' magnetic chain to check whether the ``separate'' chain can host MF or not.
``Separate'' means the presence of the vacant sites where
magnetic atoms and corresponding exchange coupling are absent.
Here, we put the number of vacant sites between each magnetic atom as $N$,
and introduce the sub-lattice denoted by A$_0$, A$_1$, $\cdots$, A$_N$.
In this sub-lattice, magnetic atoms are put on sub-lattice A$_0$.
Magnetic moments at sub-lattice A$_0$ ``rotate'' clockwise in $x$-$z$ plane with the angle $\theta$ in the similar way with ``non-separate'' case with $N=0$.
Then, the BdG Hamiltonian is given by
\begin{eqnarray}
\mathcal{H}&=&-t\sum_{i\sigma}\Bigl(
c^{\dagger}_{i\text{A$_0$}\sigma}c_{i\text{A$_1$}\sigma}
+c^{\dagger}_{i\text{A$_1$}\sigma}c_{i\text{A$_2$}\sigma}+\cdots\nonumber\\
&&+c^{\dagger}_{i\text{A$_N$}\sigma}c_{i+1,\text{A$_0$}\sigma}
+\text{H.c.}\Bigr)-\mu\sum_{ij\sigma}c^{\dagger}_{i\text{A}_j\sigma}c_{i\text{A}_j\sigma}\nonumber\\
&&
+\sum_{ij}\Bigl(\Delta c^{\dagger}_{i\text{A}_j\uparrow}c^{\dagger}_{i\text{A}_j\downarrow}+\text{H.c.}\Bigr)+\sum_{i\sigma\sigma^{\prime}}c^{\dagger}_{i\text{A}_0\sigma}(J_i)_{\sigma,\sigma^{\prime}}c_{i\text{A}_0\sigma^{\prime}}.\nonumber\\
\end{eqnarray}
Following the procedure similar to Eqs.(\ref{MF})$\sim$(\ref{PF}),
we obtain the Hamiltonian described
by $4(N+1)\times4(N+1)$ anti-symmetric matrix ${\mathcal{H}}^{\text{MF}}(k_x)$
with the basis $b_{k_x}=(b_{1\text{A}k_x\uparrow},b_{1\text{A}k_x\downarrow},\cdots,b_{2\text{A}_0k_x\uparrow},b_{2\text{A}_0k_x\downarrow}\cdots)^{T}$,
\begin{eqnarray}
\mathcal{H}=\frac{i}{4}\sum_{k_x}b_{k_x}^{\dagger}{\mathcal{H}}^{\text{MF}}(k_x)b_{k_x},\notag\\
{\mathcal{H}_{}}^{\text{MF}}(k_x)=\left( \begin{array}{cc}
{\mathcal{H}_{\;11}}^{\text{MF}}(k_x)&{\mathcal{H}_{\;12}}^{\text{MF}}(k_x)\\
{\mathcal{H}_{\;21}}^{\text{MF}}(k_x)&{\mathcal{H}_{\;22}}^{\text{MF}}(k_x)
\end{array} \right).
\end{eqnarray}
Here, the size of the matrices ${\mathcal{H}_{mn}}^{\text{MF}}(k_x) \;(m,n=1,2)$ is $2(N+1)\times2(N+1)$. 
In these matrices, only ${\mathcal{H}_{\;12}}^{\text{MF}}(k_x)$ and ${\mathcal{H}_{\;21}}^{\text{MF}}(k_x)$ have nonzero matrix elements.
\begin{equation}
{\mathcal{H}_{\;12}}^{\text{MF}}(k_x) =
\left(
\begin{array}{ccccc}
\mathcal{H}_{0} &t_{\text{hop}1}&O & \cdots &t_{\text{hop}2}(k_x)\\
t_{\text{hop}1}^{\dagger} &\mathcal{H}_{1} &t_{\text{hop}1} &\ddots &\vdots\\
O&t_{\text{hop}1}^{\dagger} &\mathcal{H}_{1} & \ddots & O \\
\vdots &\ddots & \ddots& \ddots &t_{\text{hop}1}\\
t_{\text{hop}2}^{\dagger}(k_x) & \cdots & O &t_{\text{hop}1}^{\dagger} &\mathcal{H}_{1}
\end{array}
\right),
\end{equation}
with
\begin{displaymath}
\mathcal{H}_0=\left(
\begin{array}{cc}
-\mu+J&-\Delta\\
\Delta&-\mu-J
\end{array}
\right),\;\mathcal{H}_1=
\left(
\begin{array}{cc}
-\mu&-\Delta\\
\Delta&-\mu
\end{array}
\right),
\end{displaymath}
\begin{equation*}
t_{\text{hop}1}=\left(
\begin{array}{cc}
-t&0\\
0&-t
\end{array}
\right),
\end{equation*}
\begin{equation}
\;t_{\text{hop}2}(k_x)=
\left(
\begin{array}{cc}
-t\cos(\theta/2)e^{-ik_x}&-t\sin(\theta/2)e^{-ik_x}\\
t\sin(\theta/2)e^{-ik_x}&-t\cos(\theta/2)e^{-ik_x}
\end{array}
\right).
\end{equation}
Matrix elements of ${\mathcal{H}_{\;21}}^{\text{MF}}(k_x)$ is easily obtained by the fact that this matrix is anti-symmetric.
We numerically calculate the Pfaffian of this matrix at $k_x=0$ and $k_x=\pi$ and evaluate $\mathbb{Z}_2$ topological number $\nu$. The phase diagrams of the non-collinear ``separate'' chain are shown in Figs.4 (d)$\sim$(f) for $N=1,3,$ and $10$, respectively. It is clearly seen that the non-collinear magnetic chain can be topologically non-trivial. If the interval of the magnetic atoms is much longer than the coherence length, the system cannot be topologically non-trivial. Indeed, we numerically estimate the coherence length $\xi=v_{\text{F}}/\Delta$ ($\hbar=1$) by calculating the bulk energy spectrum in the case of $N=10, J/t=1, \mu/t=-1, \Delta/t=0.1$, for instance,  (See Fig.4(f)) and get $v_{\text{F}}\approx 1$, thus, $\xi\approx 10$ which agrees well with the length of the intervals $N=10$.

\begin{figure}[h]
\begin{center}
\includegraphics[width=9.5cm]{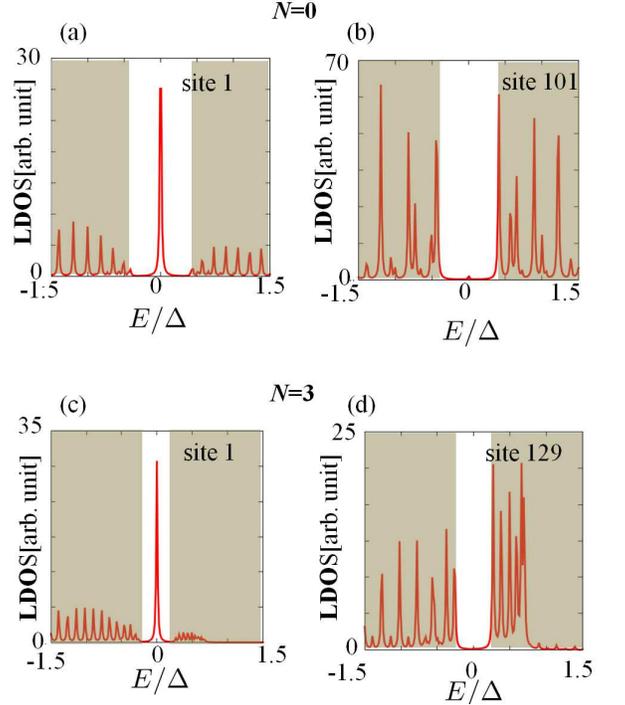}
\caption{
(a)(b)LDOS of the non-collinear and ``non-separate'' magnetic chain at site 1 and site 101. The length of the chain is 201. The shaded area represents continuum level in the system with infinite length of the chain.
(c)(d)LDOS for the ``separate'' case ($N=3$). The length of the chain is set to be 257.
The choice of the parameters is shown by filled circle in Fig.4(c) and (e).}\label{fig5}
\end{center}
\end{figure}
\begin{figure}[h]
\begin{center}
\includegraphics[width=9cm]{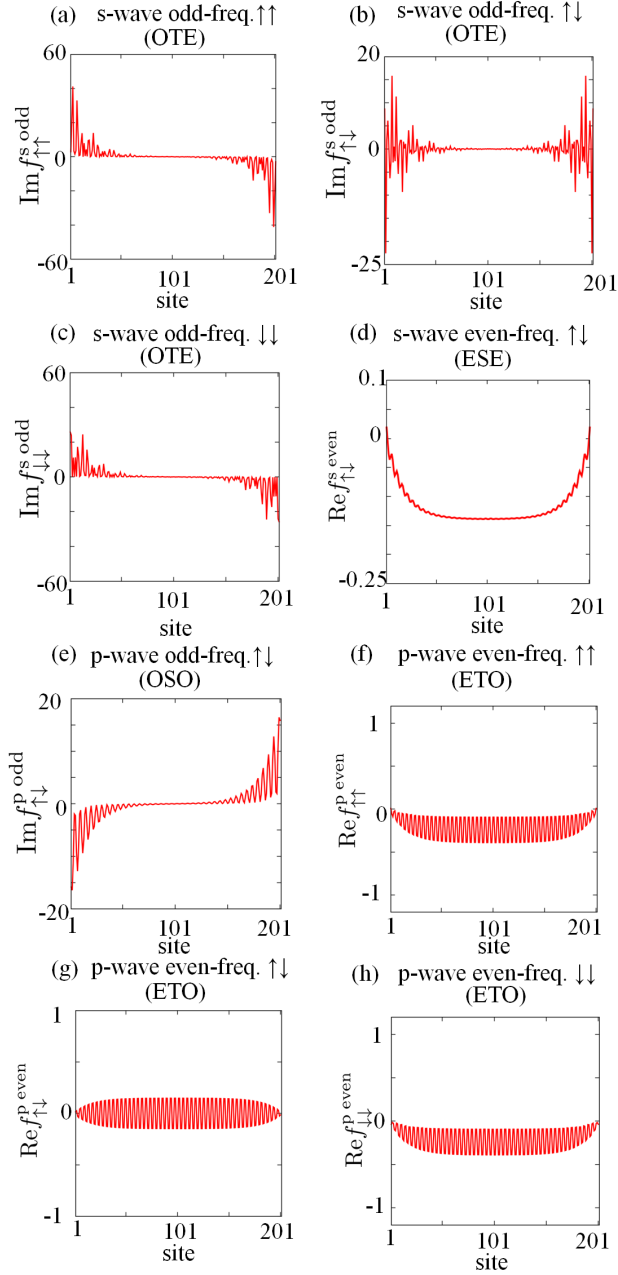}
\caption{
(a)$\sim$(h) $s$-wave odd-frequency pair, $s$-wave even-frequency pair, $p$-wave odd-frequency pair, and $p$-wave even-frequency pair of the ``non-separate'' magnetic chain.
There is no real (imaginary) value in odd-(even-) frequency pair.
The choice of the parameters is shown by filled circle in Fig.4(c).
}\label{fig6}
\end{center}
\end{figure}
\begin{figure}[h]
\begin{center}
\includegraphics[width=9cm]{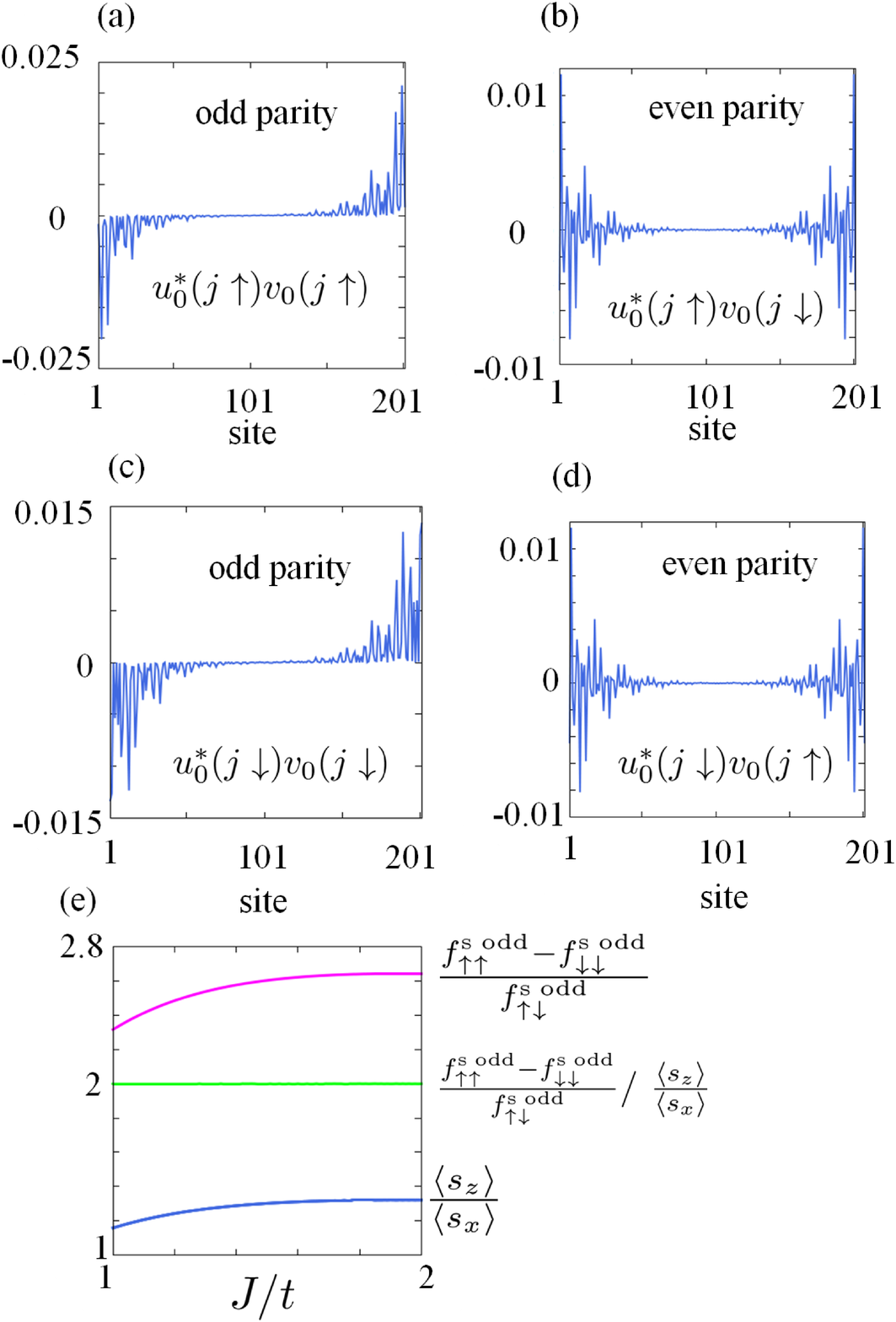}
\caption{
(a)$\sim$(d) Wave-function product of electron part $u(j\sigma)$ and hole part $v(j\sigma)$ of non-collinear magnetic chain.
(e) $\frac{f^{\text{s}\;\text{odd}}_{\uparrow\uparrow}-f^{\text{s}\;\text{odd}}_{\downarrow\downarrow}}{f^{\text{s}\;\text{odd}}_{\uparrow\downarrow}}$, $\frac{\left\langle s_z \right\rangle}{\left\langle s_x \right\rangle}$, and the ratio $\frac{f^{\text{s}\;\text{odd}}_{\uparrow\uparrow}-f^{\text{s}\;\text{odd}}_{\downarrow\downarrow}}{f^{\text{s}\;\text{odd}}_{\uparrow\downarrow}}$/ $\frac{\left\langle s_z \right\rangle}{\left\langle s_x \right\rangle}$ as a function of $J$. We fix $\mu$ as $\mu/t=-1$ (dashed arrow in Fig.4(c)) }\label{fig7}
\end{center}
\end{figure}
Next, we calculate the spatial dependence of the LDOS and pair amplitudes of the finite-size chain, which can be obtained by Eqs. (2) and (4).
In this calculation,
the direction of spin at the both edges are fixed to the $z$-direction for simplicity.
In other words, the length of the chain is set to be $N_p\times(N+1)\times l+1\;(l:\text{integer})$.
In the topologically non-trivial region,
there is zero energy Andreev bound state (ZEABS) on the edge, $i.e.$ MF exists. 
We can confirm it by the LDOS in the set of parameters denoted by filled circle in Figs.4 (c) and (e) (see Figs.5 (a) and (c)).
Similar zero energy peak is obtained in the LDOS at the edge in the case of $N=1,10$.
Figures 6(a)$\sim$(h) show pair amplitudes in the case of $N=0$. 
There exist all four types of pairings (ESE, ETO, OTE and OSO).
We note that the amplitudes of odd-frequency pair (Figs.6 (a), (b), (c), and (e)) 
are induced at the edge of the chain.
The relative signs of the amplitudes between the right and the left edges
for $\uparrow\uparrow$ and $\downarrow\downarrow$ ($\uparrow\downarrow$) components of OTE pairing are opposite (equal).
Similarly, those of OSO pairing are opposite.
Even-frequency pair amplitudes (Fig.6 (d), (f), (g), and (h)) spread the entire chain. 
In the bulk, ESE pair amplitude is constant while ETO ones show oscillating behavior in the period of $N_p$.
From these results, we can conclude that spatial dependence of the odd-frequency pairings coincide with MF.

As we have mentioned, the relative signs of odd-frequency pair amplitudes depend on their spin states.
Using Eq.(\ref{apr}), this dependence can be explained.
Below, we will focus on the case of the $s$-wave odd-frequency pair in the ``non-separate'' chain.
Nevertheless, the following discussion can be generalized for p-wave odd-frequency pairing and the odd-frequency pairings in ``separate'' chain.
We further introduce the inversion parity which is defined by the relative sign of the wave function or the pair amplitude
when we operate space inversion at the center of the chain.
The inversion parity of the $s$-wave odd-frequency pair is odd for $\uparrow\uparrow$ and $\downarrow\downarrow$ components,
whereas that is even for $\uparrow\downarrow$ component as mentioned above.
To analyze the inversion parity of the wave functions $u_0(j\sigma)$ and $v_0(j\sigma)$ in Eq.(\ref{apr}),
we numerically diagonalize the BdG Hamiltonian of the finite chain (parameters are set to be the same as the filled circle in Fig.4(c))
whose length is 201 sites.
Because of the finite size effect, the wave functions corresponding to the lowest energy are the bonding and anti-bonding states of MFs in both ends.
We use these wave functions for the actual analysis.
The results of the inversion parity are organized in Table I.
With the appropriate choice of the $U(1)$ gauge, all of the components of the wave function become real.
\begin{table}[htb]
\begin{tabular}{|c|c|c|c|c|} \hline
&$u_0(j\uparrow)$ &$u_0(j\downarrow)$ & $v_0(j\uparrow)$ &$v_0(j\downarrow)$ \\ \hline\hline
\text{parity}&\text{even}&\text{odd}&\text{odd}&\text{even}\\\hline
\end{tabular}
\caption{Inversion parity of $u_0(j\sigma)$ and $v_0(j\sigma)$. All components are real. }
\end{table}
Thus, we obtain the inversion parity of the product of $u_0^{*}(j\sigma)v_0(j\sigma)$ in Table.II.
From this table, it is expected that
the inversion parities of $f^{\text{s}\;\text{odd}}_{\uparrow\uparrow}$ and $f^{\text{s}\;\text{odd}}_{\downarrow\downarrow}$ are odd and that of $f^{\text{s}\;\text{odd}}_{\uparrow\downarrow}$ is even.
As we can see in Figs.7 (a)$\sim$(d),
this prediction based on analytical calculation meets the actual numerical results.
\begin{table}[htb]
\begin{tabular}{|c||c|c|} \hline
& $v_0j(\uparrow)$ &$v_0(j\downarrow)$ \\ \hline\hline
$u_0^{*}(j\uparrow)$&\text{odd}&\text{even}\\\hline
$u_0^{*}(j\downarrow)$&\text{even}&\text{odd}\\\hline
\end{tabular}
\caption{Inversion parity of the wave-function product $u_0^{*}v_0$ }
\end{table}
We confirm Eq.(\ref{conc}) in our set up. 
In Fig.7(e), we plot $\frac{f^{\text{s}\;\text{odd}}_{\uparrow\uparrow}-f^{\text{s}\;\text{odd}}_{\downarrow\downarrow}}{f^{\text{s}\;\text{odd}}_{\uparrow\downarrow}}$, $\frac{\left\langle s_z \right\rangle}{\left\langle s_x \right\rangle}$, and their ratio as the function of $J$ within the topologically non-trivial region shown by the dashed arrow in Fig.4(c). The ratio remains 2 as expected from Eq.(\ref{conc}). The same relation still holds at the opposite site $j=L_l$ $i.e.,$ the left side. 
Moreover, in our model calculation, the average value of $y$-component of spin of MF (Eq.(\ref{sy})) is zero, which means the direction of the spin of MF is on the $x-z$ plane.
If we set the angle of the spin of MF measured from the $x$-axis as 
$\phi=\tan^{-1}\bigl[\frac{\left\langle s_z \right\rangle}{\left\langle s_x \right\rangle}\bigr]$, Eq.(\ref{conc}) is equivalent to $\frac{f^{\text{s}\;\text{odd}}_{\uparrow\uparrow}-f^{\text{s}\;\text{odd}}_{\downarrow\downarrow}}{f^{\text{s}\;\text{odd}}_{\uparrow\downarrow}}=2\tan(\phi)$ which implies that the amplitude of the odd-frequency pair is closely related to the direction of the spin of MF.
We confirm that the relation between the inversion parity of the odd-frequency pair and that of the wave function is not affected from the choice of the $U(1)$ gauge, and Eq.(\ref{conc}) is still available.
Note also that expectation value of the spin is gauge invariant.
\begin{figure}[h]
\begin{center}
\includegraphics[width=8.5cm]{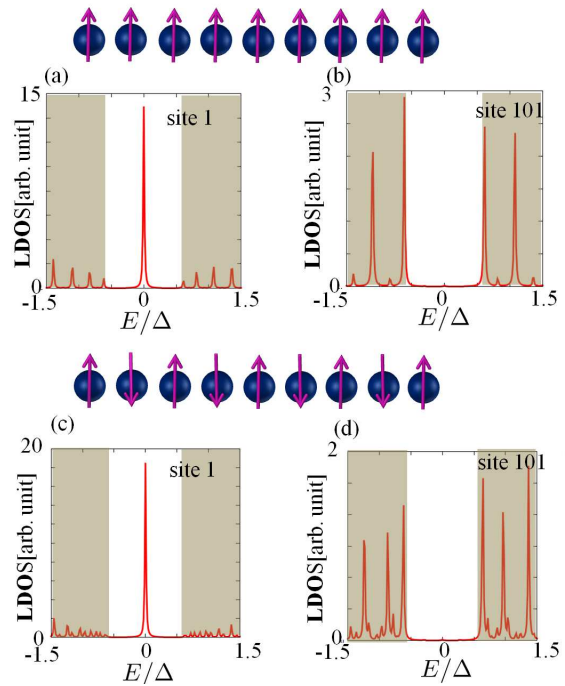}
\caption{(a)(b)((c)(d)) LDOS of the ferromagnetic (anti-ferromagnetic) chain at site 1 and site 101. The length of the chain is 201. Parameters are set as $\mu/t=-1, \Delta/t=0.1, J/t=1.5$, and $\lambda_R/t=1.0$. The shaded area represents continuum level in the system with infinite length of the chain.}\label{fig8}
\end{center}
\end{figure}
\section{Effect of the Rashba-type SOC on magnetic chains}\label{sec5}
In this section, we analyze the magnetic chain on spin singlet $s$-wave superconductor with SOC especially for ferromagnetic and anti-ferromagnetic configurations. As we mentioned in the previous section, ferromagnetic and anti-ferromagnetic chain 
cannot be topologically non-trivial and Majorana fermion is absent at the edges. With SOC, however, both of the cases can be topologically non-trivial. We can obtain the model Hamiltonian with SOC only by replacing the hopping term in Eq.(\ref{model2}) with 
\begin{equation}
\sum_{i\sigma\sigma^{\prime}}c^{\dagger}_{i+1\sigma}\left( \begin{array}{cc}
-t & \lambda_R/2 \\
-\lambda_R/2& -t 
\end{array} \right)c_{i\sigma^{\prime}}+\text{H.c.},
\end{equation}
where $\lambda_R$ represents Rashba-type SOC. 
We set $\theta=0$ for ferromagnetic, $\theta=\pi$ for anti-ferromagnetic case in Eq.(\ref{model2}). 
After the similar procedure shown in Eqs.(\ref{gauge})$\sim$(\ref{nontrivial}), we get topologically non-trivial condition: 
\begin{eqnarray}
\sqrt{(|2t\cos(\theta/2)+\lambda_R\sin(\theta/2)|-|\mu|)^2+\Delta^2}<|J|\notag\\<\sqrt{(|
2t\cos(\theta/2)+\lambda_R\sin(\theta/2)|+|\mu|)^2+\Delta^2}\label{cond}.
\end{eqnarray}
For $\theta=0$, the condition is not changed from Eq.(\ref{nontrivial}), but we confirm that the bulk energy gap is opened by SOC. Thus, we can make the ferromagnetic chain non-trivial. 
Actually, we can see the ZEABS localized on the edge as shown in Figs.8(a) and (b).
Recently, this ferromagnetic chain has great attention since the experiment by Nadj-Perge et al. 
has confirmed the zero energy peak of the LDOS at the edge, which supports the presence of MF.\cite{yazdaniex}
The model of this ferromagnetic chain with SOC is essentially equivalent to that of a semiconducting nanowire deposited on $s$-wave superconductor with Zeeman field.\cite{alicea10} 
In the case of anti-ferromagnetic chain, $i.e.$ $\theta=\pi$, inequality (\ref{cond}) becomes $\sqrt{(|\lambda_R|-|\mu|)^2+\Delta^2}<|J|<\sqrt{(|\lambda_R|+|\mu|)^2+\Delta^2}$, thus, with non-zero $\lambda_R$, the left side and the right side in inequality (\ref{cond}) are never equal to be the same value, which implies anti-ferromagnetic chain with SOC can be topologically non-trivial. We plot the LDOS in Figs.8 (c) and (d). We clearly see that there is ZEABS $i.e.$ MF on the edge of the chain. \par
We also calculate the odd-, and even-frequency pair amplitudes in Figs.9(a)$\sim$(h) and Figs.10(a)$\sim$(h) for ferromagnetic and anti-ferromagnetic cases respectively.
As shown in Appendix C, 
OTE $\uparrow\downarrow$ pair amplitude is allowed to exist in the infinite system in the ferromagnetic configuration.
On the other hand, the other types of odd-frequency pair amplitudes
are found to exist on the edge as shown in Figs.9 (a)$\sim$(c), and (e). Similarly, all types of the odd-frequency pair amplitude are generated on the edge in Figs.10 (a)$\sim$(c), and (e) in the anti-ferromagnetic configuration. 
By contrast to the case in ferromagnetic and anti-ferromagnetic chains without SOC,
odd-frequency pairings which are absent in the infinite chain develop at the edge accompanied by the emergence of MF. 
In addition, the inversion parity of the
odd-frequency pair amplitudes are completely the same as that in the case of non-collinear array of magnetic atoms as discussed in the previous section for both ferromagnetic and anti-ferromagnetic chains.
\\
\begin{figure}[h]
\begin{center}
\includegraphics[width=8.5cm]{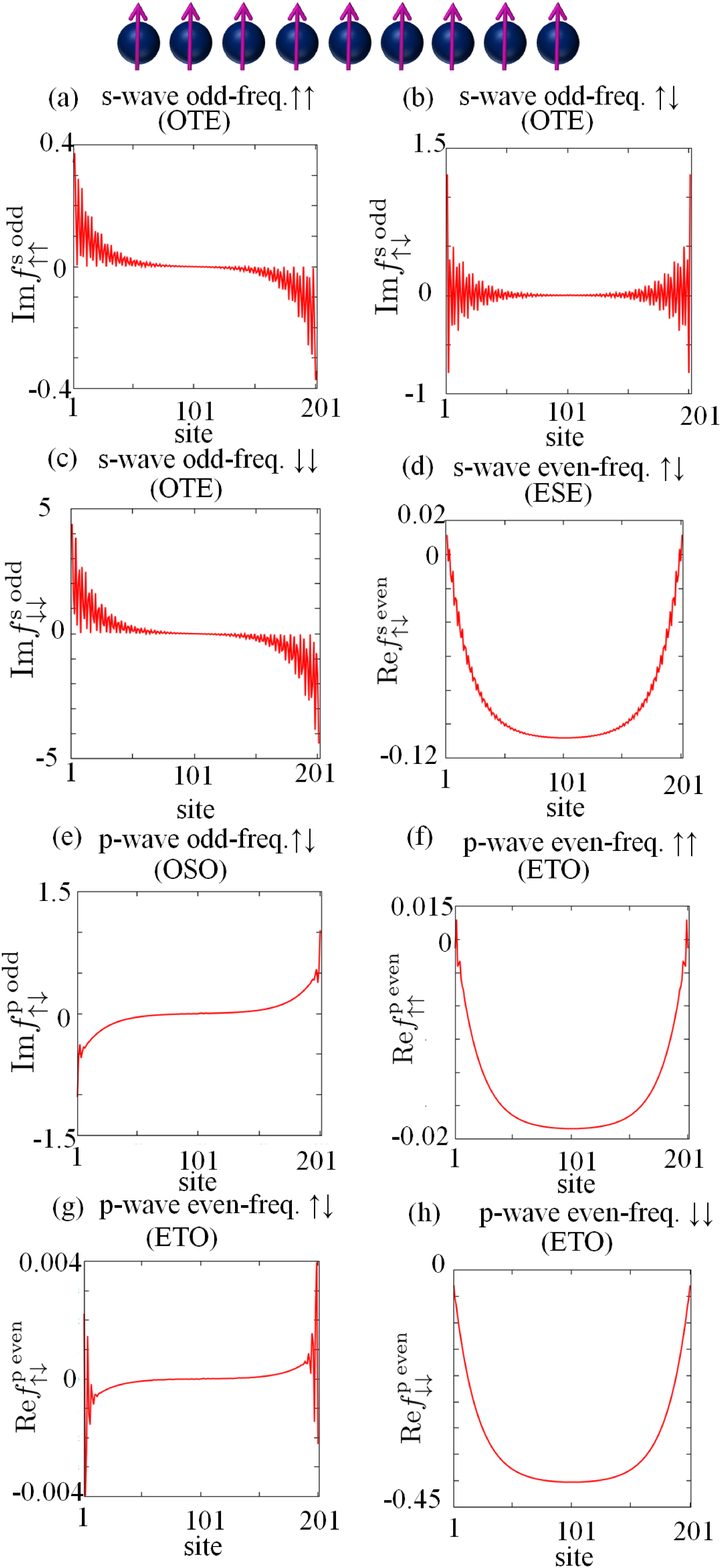}
\caption{(a)$\sim$(h) Pair amplitudes of $s$-wave odd-frequency, $s$-wave even-frequency, $p$-wave odd-frequency and $p$-wave even-frequency in the ferromagnetic chain.
Parameters are set to be the same as Fig.8.}\label{fig9}
\end{center}
\end{figure}
\begin{figure}[h]
\begin{center}
\includegraphics[width=8.5cm]{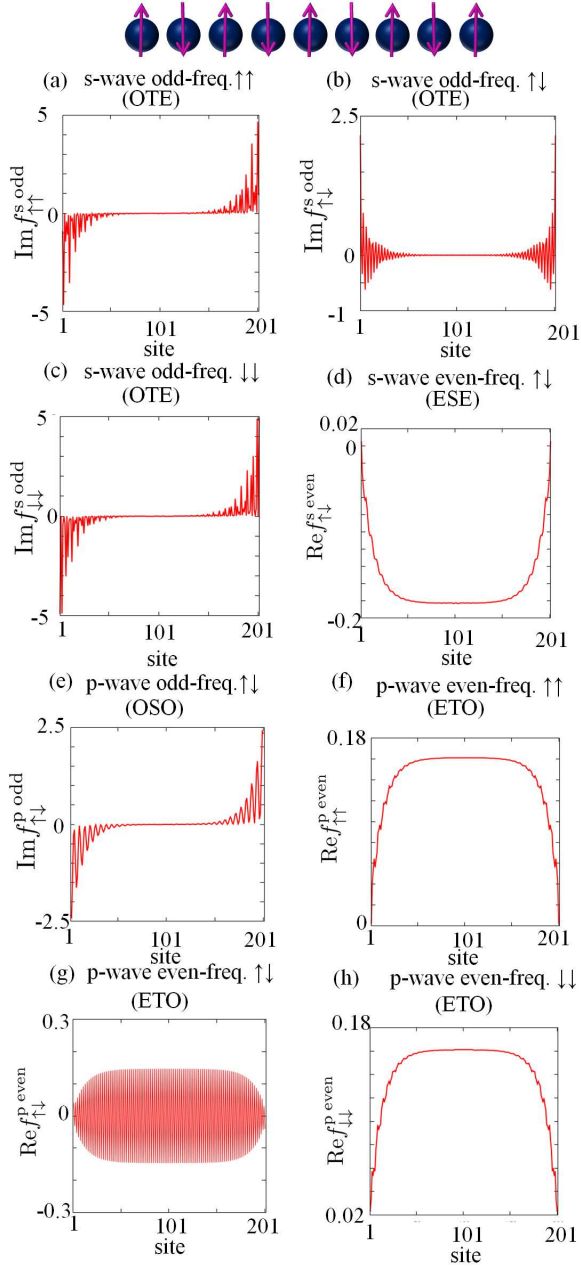}
\caption{
(a)$\sim$(h) Pair amplitudes $s$-wave odd-frequency, $s$-wave even-frequency, $p$-wave odd-frequency and $p$-wave even-frequency in the anti-ferromagnetic chain.
Parameters are set to be the same as the calculation in Fig.8.}\label{fig10}
\end{center}
\end{figure}
\section{Conclusion}\label{sec6}
In this paper, the pair amplitude of the Cooper pair with ESE, ETO, OTE and OSO symmetries, and the LDOS 
have been studied in various types of magnetic chain on spin-singlet $s$-wave superconductor.
It is clarified that odd-frequency pairings are generated at the edge when the 1D chain of non-collinear magnetic atoms without SOC on superconductor is topologically non-trivial. We also reveal that the spatial dependence 
of $s$-wave OTE and $p$-wave OSO pair amplitudes at the edge can be explained 
by the inversion parity of the wave function and the direction of the spin of MF. Even when the magnetic atoms are positioned at the intervals, the chain can be topologically non-trivial. 
In the presence of SOC, ferromagnetic and anti-ferromagnetic chain can be topologically non-trivial and we obtain the similar correspondence between the odd-frequency pair amplitude and MF. 
Due to the translational and spin-rotational symmetry breaking, three types of paring symmetries -- OTE, OSO, and ETO,
which are absent in the bulk spin-singlet $s$-wave superconductor, are induced. 
It can be summarized that 
an array of magnetic atoms on conventional superconductor is an exotic and 
intriguing system not only in the topological perspective, but also from the view point of the symmetries of the Cooper pair. 
Based on the obtained results in this paper, 
we can say that 
the detection of the zero energy peak of 
the LDOS by scanning tunneling microscopy at the edge of magnetic chain
is a strong evidence for the existence of odd-frequency pairing. 
The relation between the zero energy LDOS and odd-frequency pairing 
have been clarified in several different systems \cite{odd3,Yokoyamavortex,
Higashitani}.
For this reason, we can say that zero energy peak of LDOS 
relates directly with 
the presence of odd-frequency pair.
In this paper, main results are obtained based on BdG equation with mean field approximation. 
It is a standard method to study inhomogeneous superconductor. 
Since the inducement of the odd-frequency pairing is based on the 
broken symmetry of the Hamiltonian,   
our obtained results are not changed essentially even if we go beyond 
mean field approximation. 
\par

To close this section, we suggest another possible experiment to detect odd-frequency pairing in an array of magnetic chain on spin-singlet $s$-wave superconductor, where topologically non-trivial state is realized. 
If odd-frequency pair amplitude is enhanced at the edge of the magnetic chain,
the symmetry of pair amplitude at the edge and in the bulk are different.
We can detect this difference by the local Josephson 
current \cite{Yokoyamavortex} by using superconducting STM tip.
When conventional superconducting tip (ESE pairing) contacts the edge of the chain or the bulk of superconductor,
different current phase relation of Josephson current is expected due to the difference of the pairing symmetry. 
The former case corresponds to  i)ESE/(OSO +OTE) junction
and the latter one corresponds to ii)ESE/ESE junction (See Fig.11). 
In the case i), the first order Josephson coupling is absent, 
while it exists in the case ii).
Furthermore, if the electronic state at the protrusion of the tip is magnetic, which is realized by putting Fe atom at the tip, for instance,
the pairing symmetry of the tip includes the odd-frequency and spin-triplet 
pairing. 
Thus, STM tip can measure more details of spin state in pairing amplitude on the magnetic chain.

\begin{figure}[h]
\begin{center}
\includegraphics[width=6cm]{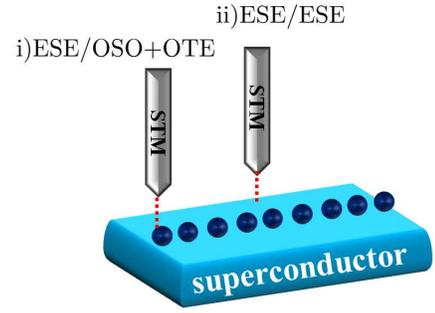}
\caption{A schematic picture of the experiment of detecting odd-frequency pair. Dots on the superconductor represent magnetic atoms}\label{fig11}
\end{center}
\end{figure}

\section*{Acknowledgments}
This work was supported in part by a Grant-in Aid
for Scientific Research from MEXT of Japan, ``Topological
Quantum Phenomena, Grants No. 22103005, No. 20654030
and No. 22540383. 
\appendix
\section{Matsubara Green's function of the infinite ferromagnetic chain without SOC}
In this appendix, we analyze symmetries of the pair amplitudes 
of the infinite ferromagnetic chain without SOC, where only the spin-rotational symmetry is broken. 
Model Hamiltonian is given by
\begin{eqnarray}
\mathcal{H}=-t\sum_{i \sigma}\Bigl(c^{\dagger}_{i\sigma}c_{i+1\sigma}+\text{H.c.}\Bigr)
-\mu\sum_{i,\sigma}c^{\dagger}_{i\sigma}c_{i\sigma}\notag\\
+\sum_i\Bigl( \Delta c^{\dagger}_{i\uparrow}c^{\dagger}_{i\downarrow}
+\text{H.c.}\Bigr)
+J\sum_{i\sigma,\sigma^{\prime}}c^{\dagger}_{i\sigma}(\sigma_z)_{\sigma,\sigma^{\prime}}c_{i\sigma^{\prime}}, \label{model-fro}
\end{eqnarray}
by  Fourier transformation
\begin{equation}
c_{j\sigma}=\sum_{k_x}c_{k_x\sigma}e^{-ijk_x}.
\end{equation}
We obtain the $4\times4$ BdG Hamiltonian with the basis $c_{k_x}=(c_{k_x\uparrow},c_{k_x\downarrow},c^{\dagger}_{-k_x\uparrow},c^{\dagger}_{-k_x\downarrow})^{\text{T}}$:

\begin{equation}
\mathcal{H}=\sum_{k_x}c_{k_x}^{\dagger}\mathcal{H}(k_x)c_{k_x},
\end{equation}
\begin{equation}
\mathcal{H}(k_x)=\left( \begin{array}{cccc}
\xi_{k_x}+J&0&0&\Delta\\
0&\xi_{k_x}-J&-\Delta&0\\
0&-\Delta&-\xi_{k_x}-J&0\\
\Delta&0&0&-\xi_{k_x}+J
\end{array} \right),
\end{equation}
with 
\begin{equation}
\xi_{k_x}=-2t\cos k_x-\mu\label{xi}.
\end{equation}
Above Hamiltonian can be decomposed into two $2\times2$ matrices, that is, $\mathcal{H}=\mathcal{H}_1 \bigoplus \mathcal{H}_2$:
\begin{eqnarray}
\mathcal{H}_1=&&\frac{1}{2}\sum_{k_x}c_{1k_x}^{\dagger}\mathcal{H}_1(k_x)c_{1k_x},\label{bdg22}\\
\mathcal{H}_1(k_x)=&&\left( \begin{array}{cc}
\xi_{k_x}+J&\Delta\\
\Delta&-\xi_{k_x}+J\\
\end{array} \right)\notag,
\end{eqnarray}
with the basis $c_{1k_x}=(c_{k_x\uparrow},c^{\dagger}_{-k_x\downarrow})$ and 
\begin{eqnarray}
\mathcal{H}_2=&&\frac{1}{2}\sum_{k_x}c_{2k_x}^{\dagger}\mathcal{H}_2(k_x)c_{2k_x},\label{bdg22}\\
\mathcal{H}_2(k_x)=&&\left( \begin{array}{cc}
\xi_{k_x}-J&-\Delta\\
-\Delta&-\xi_{k_x}-J\\
\end{array} \right)\notag,
\end{eqnarray}
with the basis $c_{2k_x}=(c_{k_x\downarrow},c^{\dagger}_{-k_x\uparrow})$.
Matsubara Green's function is defined by
\begin{eqnarray}
G_1(\omega_n,k_x)=\frac{1}{i\omega-\mathcal{H}_1(k_x)}=\left( \begin{array}{cc}
g_{\uparrow\uparrow}&f_{\uparrow\downarrow}\\
\tilde{f}_{\downarrow\uparrow}&\tilde{g}_{\downarrow\downarrow}\\
\end{array} \right),\label{green3}\\
G_2(\omega_n,k_x)=\frac{1}{i\omega-\mathcal{H}_2(k_x)}=\left( \begin{array}{cc}
g_{\downarrow\downarrow}&f_{\downarrow\uparrow}\\
\tilde{f}_{\uparrow\downarrow}&\tilde{g}_{\uparrow\uparrow}\\
\end{array} \right).
\label{green4}
\end{eqnarray}
They are calculated to be
\begin{equation}
G_1(\omega_n,k_x)=\left( \begin{array}{cc}
\frac{i\omega_n-J+\xi_{k_x}}{\Pi(k_x,\omega_n)-2i\omega_nJ}&\frac{\Delta}{\Pi(k_x,\omega_n)-2i\omega_nJ}\\
\frac{\Delta}{\Pi(k_x,\omega_n)-2i\omega_nJ}&\frac{i\omega_n-J-\xi_{k_x}}{\Pi(k_x,\omega_n)-2i\omega_nJ}\\
\end{array} \right),
\end{equation}
\begin{equation}
G_2(\omega_n,k_x)=\left( \begin{array}{cc}
\frac{i\omega_n+J+\xi_{k_x}}{\Pi(k_x,\omega_n)+2i\omega_nJ}&\frac{\Delta}{\Pi(k_x,\omega_n)+2i\omega_nJ}\\
\frac{\Delta}{\Pi(k_x,\omega_n)+2i\omega_nJ}&\frac{i\omega_n+J-\xi_{k_x}}{\Pi(k_x,\omega_n)+2i\omega_nJ}\\
\end{array} \right),
\end{equation}
with
\begin{equation}
\Pi(k_x,\omega_n)=-\Delta^2-\xi_{k_x}^2+J^2-\omega_n^2.\label{pi}
\end{equation}
Therefore, the anomalous part of Matsubara Green's functions, $i.e.$, pair amplitudes $f_{\uparrow\downarrow}$ and $f_{\downarrow\uparrow}$, are given by 
\begin{equation}
f_{\uparrow\downarrow}=\frac{\Delta}{\Pi(k_x,\omega_n)-2i\omega_nJ}\;\;f_{\downarrow\uparrow}=\frac{-\Delta}{\Pi(k_x,\omega_n)+2i\omega_nJ}.\label{anomalous}
\end{equation}
Then, we can obtain the spin-singlet and spin-triplet components of 
pair amplitudes 
\begin{equation}
\frac{f_{\uparrow\downarrow}-f_{\downarrow\uparrow}}{2}=\frac{\Delta\Pi(k_x,\omega_n)}{\Pi(k_x,\omega_n)^2+4\omega_n^2J^2}\label{singlet}
\end{equation}
\begin{equation}
\frac{f_{\uparrow\downarrow}+f_{\downarrow\uparrow}}{2}=\frac{2i\Delta\omega_nJ}{\Pi(k_x,\omega_n)^2+4\omega_n^2J^2}.\label{triplet}
\end{equation}
Here, $\uparrow\uparrow$ and $\downarrow\downarrow$ components of the spin-triplet pair amplitudes are absent.
Based on Eqs.(\ref{xi}), (\ref{pi}), (\ref{singlet}), and (\ref{triplet}), 
the symmetry of the spin-singlet pair amplitude is even-frequency and even-parity
because it is not changed under the inversion operation ($\omega_n\to-\omega_n$, $k_x\to-k_x$).
On the other hand, that of spin-triplet pair amplitude is odd-frequency and even parity.
Obtained symmetries are summarized in Table III. These results are consistent with the numerically obtained results 
in the main text. 
(see, Figs.2(d)(e))
\begin{table}[htb]
\begin{tabular}{|c|c|c|c|c|c|c|c|c|c|c|c|} \hline
\multicolumn{3}{|c|}{\text{OTE} }& \multicolumn{3}{|c|}{\text{ESE} }& \multicolumn{3}{|c|}{\text{OSO} }& \multicolumn{3}{|c|}{\text{ETO} }\\ \hline\hline
$\uparrow\uparrow$&$\uparrow\downarrow$&$\downarrow\downarrow$& \multicolumn{3}{|c|}{$\uparrow\downarrow$}&\multicolumn{3}{|c|}{$\uparrow\downarrow$}&$\uparrow\uparrow$&$\uparrow\downarrow$&$\downarrow\downarrow$\\\hline
$\times$&$\bigcirc$&$\times$& \multicolumn{3}{|c|}{$\bigcirc$}&\multicolumn{3}{|c|}{$\times$}&$\times$&$\times$&$\times$\\\hline
\end{tabular}
\caption{Symmetries of the pair amplitudes of the infinite ferromagnetic chain without SOC. The arrows in the second line denotes the spin configuration of Cooper pair. There are three configurations for spin-triplet and one for spin-singlet pairing.
Pair amplitudes marked by $\bigcirc$ ($\times$) are present (absent).}
\end{table}

\section{Matsubara Green's function of the infinite 
anti-ferromagnetic chain without SOC}
In appendix B, we investigate the symmetries of the pair amplitudes of the infinite anti-ferromagnetic chain without SOC.
In the model of the anti-ferromagnetic chain, there are two sites in the unit cell denoted by A and B. 
The direction of the spin is chosen to be $+z$ ($-z$) direction at A (B) site.
The Hamiltonian is given by
\begin{eqnarray}
\mathcal{H}=-t\sum_{i \sigma}\Bigl(c^{\dagger}_{i\text{A}\sigma}c_{i\text{B}\sigma}+c^{\dagger}_{i+1\text{B}\sigma}c_{i\text{A}\sigma}+\text{H.c.}\Bigr)
\notag\\
-\mu\sum_{i,\sigma}c^{\dagger}_{i\sigma}c_{i\sigma}
+\sum_i\Bigl( \Delta c^{\dagger}_{i\uparrow}c^{\dagger}_{i\downarrow}+\text{H.c.}\Bigr)\notag\\
+J\sum_{i\sigma,\sigma^{\prime}}\Bigl(c^{\dagger}_{i\text{A}\sigma}(\sigma_z)_{\sigma,\sigma^{\prime}}c_{i\text{A}\sigma^{\prime}}-c^{\dagger}_{i\text{B}\sigma}(\sigma_z)_{\sigma,\sigma^{\prime}}c_{i\text{B}\sigma^{\prime}}\Bigr)\label{model-apn}.
\end{eqnarray}
If we perform Fourier transform as
\begin{eqnarray}
c_{j\text{A}\sigma}&=&\sum_{k_x}c_{k_x\text{A}\sigma}e^{-ijk_x}\notag\\
c_{j\text{B}\sigma}&=&\sum_{k_x}c_{k_x\text{B}\sigma}e^{-i(j+1)k_x}, 
\end{eqnarray}
we obtain the decomposed $4\times4$ BdG Hamiltonian 
by similar way in  Appendix A. $\mathcal{H}=\mathcal{H}_1 \bigoplus \mathcal{H}_2$,
with
\begin{equation}
\mathcal{H}_1=\frac{1}{2}\sum_{k_x}c_{1k_x}^{\dagger}\mathcal{H}_1(k_x)c_{1k_x},
\end{equation}
\begin{equation}
\mathcal{H}_1(k_x)=\left( \begin{array}{cccc}
J-\mu&-2t\cos k_x&\Delta&0\\
-2t\cos k_x&-J-\mu&0&\Delta\\
\Delta&0&J+\mu&2t\cos k_x\\
0&\Delta&2t\cos k_x&-J+\mu
\end{array} \right)
\end{equation}
with the basis $c_{1k_x}=(c_{k_x\text{A}\uparrow},c_{k_x\text{B}\uparrow},c^{\dagger}_{-k_x\text{A}\downarrow},c^{\dagger}_{-k_x\text{B}\downarrow})^{\text{T}}$ and
\begin{equation}
\mathcal{H}_2=\frac{1}{2}\sum_{k_x}c_{2k_x}^{\dagger}\mathcal{H}_2(k_x)c_{2k_x},
\end{equation}
\begin{equation}
\mathcal{H}_2(k_x)=\left( \begin{array}{cccc}
-J-\mu&-2t\cos k_x&-\Delta&0\\
-2t\cos k_x&J-\mu&0&-\Delta\\
-\Delta&0&-J+\mu&2t\cos k_x\\
0&-\Delta&2t\cos k_x&J+\mu
\end{array} \right)
\end{equation}
with the basis $c_{2k_x}=(c_{k_x\text{A}\downarrow},c_{k_x\text{B}\downarrow},c^{\dagger}_{-k_x\text{A}\uparrow},c^{\dagger}_{-k_x\text{B}\uparrow})^{\text{T}}$.
Matsubara Green's function is defined as
\begin{equation}
G_1(\omega_n,k_x)=\frac{1}{i\omega-\mathcal{H}_1(k_x)}=\left( \begin{array}{cc}
G_{\uparrow\uparrow}&F_{\uparrow\downarrow}\\
\tilde{F}_{\downarrow\uparrow}&\tilde{G}_{\downarrow\downarrow}\\
\end{array} \right)\label{green2}.
\end{equation}
\begin{equation}
G_2(\omega_n,k_x)=\frac{1}{i\omega-\mathcal{H}_2(k_x)}=\left( \begin{array}{cc}
G_{\downarrow\downarrow}&F_{\downarrow\uparrow}\\
\tilde{F}_{\uparrow\downarrow}&\tilde{G}_{\uparrow\uparrow}\\
\end{array} \right)\label{green44}.
\end{equation}
The size of the matrices in the right side in Eqs. (\ref{green2}) and (\ref{green44}) is $4\times4$.
Especially, we focus on the anomalous part $F_{\uparrow\downarrow}$ and $F_{\downarrow\uparrow}$:
\begin{equation}
F_{\uparrow\downarrow}=\left( \begin{array}{cc}
F^{\text{AA}}_{\uparrow\downarrow}&F^{\text{AB}}_{\uparrow\downarrow}\\
F^{\text{BA}}_{\uparrow\downarrow}&F^{\text{BB}}_{\uparrow\downarrow}\\
\end{array} \right),  
\end{equation}
\begin{equation}
F_{\downarrow\uparrow}=\left( \begin{array}{cc}
F^{\text{AA}}_{\downarrow\uparrow}&F^{\text{AB}}_{\downarrow\uparrow}\\
F^{\text{BA}}_{\downarrow\uparrow}&F^{\text{BB}}_{\downarrow\uparrow}\\
\end{array} \right). 
\end{equation}
It is noted that the symmetries of $F^{\text{AA}}_{\uparrow\downarrow}$ and $F^{\text{BB}}_{\uparrow\downarrow}$ are even parity while those of $F^{\text{AB}}_{\uparrow\downarrow}$ and $F^{\text{BA}}_{\uparrow\downarrow}$ are odd parity. This is because 
$F^{\text{AA}}_{\uparrow\downarrow}$ and $F^{\text{BB}}_{\uparrow\downarrow}$ correspond to the on-site pair amplitude while $F^{\text{AB}}_{\uparrow\downarrow}$ and $F^{\text{BA}}_{\uparrow\downarrow}$ do to the one between adjacent sites. 
When we calculate the Matsubara Green's function, we use the following formula
\begin{equation}
\left( \begin{array}{cc}
X&Y\\
Z&W\\
\end{array} \right)^{-1}=\left( \begin{array}{cc}
(X-YW^{-1}Z)^{-1}&(Z-WY^{-1}X)^{-1}\\
(Y-XZ^{-1}W)^{-1}&(W-ZX^{-1}Y)^{-1}\\
\end{array} \right)\label{formula},
\end{equation}
where the size of the matrix $X, Y, Z,$ and $W$ is the same. From Eqs.(\ref{green2}) $\sim$ (\ref{formula}), we obtain $F_{\uparrow\downarrow}$ and $F_{\downarrow\uparrow}$. For instance, if one wants to find $F_{\uparrow\downarrow}$, 
$X, Y, Z,$ and $W$ are given by
\begin{eqnarray}
X=\left( \begin{array}{cc}
i\omega_n-J+\mu&2t\cos k_x\\
2t\cos k_x&i\omega_n+J+\mu\\
\end{array} \right), \notag\\Y=Z=\left( \begin{array}{cc}
-\Delta&0\\
0&-\Delta\\
\end{array} \right),\notag\\
W=\left(\begin{array}{cc}
i\omega_n-J-\mu&-2t\cos k_x\\
-2t\cos k_x&i\omega_n+J-\mu\\
\end{array} \right).
\end{eqnarray}
After straightforward calculation, 
\begin{equation}
F_{\uparrow\downarrow}=\frac{-\Delta}{\beta(\omega_n,k_x)}\left( \begin{array}{cc}
\alpha(\omega_n,k_x)-2i\omega_nJ&-4t(J+\mu)\cos k_x\\
4t(J-\mu)\cos k_x&\alpha(\omega_n,k_x)+2i\omega_nJ\\
\end{array} \right),
\end{equation}
with 
\begin{eqnarray}
\alpha(\omega_n,k_x)&=&\omega_n^2+\Delta^2+\mu^2+4t^2\cos^2k_x-J^2\label{alpha}\\
\beta(\omega_n,k_x)&=&\alpha^2(\omega_n,k_x)+4\omega_n^2J^2\notag\\
&+&16t^2(J^2-\mu^2)\cos^2k_x\label{beta}.
\end{eqnarray}
Similarly, 
\begin{equation}
F_{\downarrow\uparrow}=\frac{\Delta}{\beta(\omega_n,k_x)}\left( \begin{array}{cc}
\alpha(\omega_n,k_x)+2i\omega_nJ&4t(J-\mu)\cos k_x\\
-4t(J+\mu)\cos k_x&\alpha(\omega_n,k_x)-2i\omega_nJ\\
\end{array} \right).
\end{equation}
Then, spin-singlet and  spin-triplet components of pair amplitudes are
\begin{eqnarray}
\frac{1}{2}(F_{\uparrow\downarrow}-F_{\downarrow\uparrow})=\left( \begin{array}{cc}
\frac{F^{\text{AA}}_{\uparrow\downarrow}-F^{\text{AA}}_{\downarrow\uparrow}}{2}&\frac{F^{\text{AB}}_{\uparrow\downarrow}-F^{\text{AB}}_{\downarrow\uparrow}}{2}\\
\frac{F^{\text{BA}}_{\uparrow\downarrow}-F^{\text{BA}}_{\downarrow\uparrow}}{2}&\frac{F^{\text{BB}}_{\uparrow\downarrow}-F^{\text{BB}}_{\downarrow\uparrow}}{2}\\
\end{array} \right)&&\notag\\
=\left( \begin{array}{cc}
\frac{-\Delta\alpha(\omega_n,k_x)}{\beta(\omega_n,k_x)}&\frac{4\Delta t\mu\cos k_x}{\beta(\omega_n,k_x)}\\
\frac{4\Delta t\mu\cos k_x}{\beta(\omega_n,k_x)}&\frac{-\Delta\alpha(\omega_n,k_x)}{\beta(\omega_n,k_x)}\\
\end{array} \right),&&\label{il}
\end{eqnarray}

\begin{eqnarray}
\frac{1}{2}(F_{\uparrow\downarrow}+F_{\downarrow\uparrow})=\left( \begin{array}{cc}
\frac{F^{\text{AA}}_{\uparrow\downarrow}+F^{\text{AA}}_{\downarrow\uparrow}}{2}&\frac{F^{\text{AB}}_{\uparrow\downarrow}+F^{\text{AB}}_{\downarrow\uparrow}}{2}\\
\frac{F^{\text{BA}}_{\uparrow\downarrow}+F^{\text{BA}}_{\downarrow\uparrow}}{2}&\frac{F^{\text{BB}}_{\uparrow\downarrow}+F^{\text{BB}}_{\downarrow\uparrow}}{2}\\
\end{array} \right)&&\notag\\
=\left( \begin{array}{cc}
\frac{2i\Delta\omega_nJ}{\beta(\omega_n,k_x)}&\frac{4\Delta tJ\cos k_x}{\beta(\omega_n,k_x)}\\
\frac{-4\Delta tJ\cos k_x}{\beta(\omega_n,k_x)}&\frac{-2i\Delta\omega_nJ}{\beta(\omega_n,k_x)}\\
\end{array} \right).&&\label{li}
\end{eqnarray}
\begin{table}[htb]
\begin{tabular}{|c|c|c|c|c|c|c|c|c|c|c|c|} \hline
\multicolumn{3}{|c|}{\text{OTE} }& \multicolumn{3}{|c|}{\text{ESE} }& \multicolumn{3}{|c|}{\text{OSO} }& \multicolumn{3}{|c|}{\text{ETO} }\\ \hline\hline
$\uparrow\uparrow$&$\uparrow\downarrow$&$\downarrow\downarrow$& \multicolumn{3}{|c|}{$\uparrow\downarrow$}&\multicolumn{3}{|c|}{$\uparrow\downarrow$}&$\uparrow\uparrow$&$\uparrow\downarrow$&$\downarrow\downarrow$\\\hline
$\times$&$\bigcirc$&$\times$& \multicolumn{3}{|c|}{$\bigcirc$}&\multicolumn{3}{|c|}{$\times$}&$\times$&$\bigcirc$&$\times$\\\hline
\end{tabular}
\caption{Symmetries of the pair amplitude of the infinite anti-ferromagnetic chain without SOC similar to Table III.}
\end{table}
For spin-singlet pair amplitude, only ESE and OSO are allowed 
in consistent with Fermi-Dirac statistics.
However, from Eqs.(\ref{alpha}), (\ref{beta}), and (\ref{il}), the symmetries of the singlet pair amplitudes, $\frac{F^{\text{AA}}_{\uparrow\downarrow}-F^{\text{AA}}_{\downarrow\uparrow}}{2}$ and $\frac{F^{\text{BB}}_{\uparrow\downarrow}-F^{\text{BB}}_{\downarrow\uparrow}}{2}$ ($\frac{F^{\text{AB}}_{\uparrow\downarrow}-F^{\text{AB}}_{\downarrow\uparrow}}{2}$ and $\frac{F^{\text{BA}}_{\uparrow\downarrow}-F^{\text{BA}}_{\downarrow\uparrow}}{2}$), which are even-parity (odd-parity), are even-frequency because they do not change sign when we operate $\omega_n\to-\omega_n$. Thus, only ESE symmetry is possible in the spin-singlet sector.
From Eqs.(\ref{alpha}), (\ref{beta}), and (\ref{li}), the symmetries of spin-triplet even-parity pair amplitude, $\frac{F^{\text{AA}}_{\uparrow\downarrow}+F^{\text{AA}}_{\downarrow\uparrow}}{2}$ and $\frac{F^{\text{BB}}_{\uparrow\downarrow}+F^{\text{BB}}_{\downarrow\uparrow}}{2}$, are odd-frequency because their sings are changed by the operation $\omega_n\to-\omega_n$. On the other hand, spin-triplet odd-parity pair amplitude, $\frac{F^{\text{AB}}_{\uparrow\downarrow}+F^{\text{AB}}_{\downarrow\uparrow}}{2}$ and $\frac{F^{\text{BA}}_{\uparrow\downarrow}+F^{\text{BA}}_{\downarrow\uparrow}}{2}$, are even-frequency because they have only the quadratic term of $\omega_n$. We summarize these results in Table IV.
They are consistent with the numerically obtained results in the main text (Figs.3(d)$\sim$(f)). 

\section{Matsubara Green's function of the infinite ferromagnetic chain with Rashba-type SOC}
In appendix C, we study the symmetries of the infinite ferromagnetic chain with Rashba-type SOC. As explained in the main text, the model Hamiltonian is given by
\begin{eqnarray}
\mathcal{H}=\sum_{i \sigma}\Bigl(c^{\dagger}_{i+1\sigma}Tc_{i\sigma}+\text{H.c.}\Bigr)
-\mu\sum_{i,\sigma}c^{\dagger}_{i\sigma}c_{i\sigma}\notag\\
+\sum_i\Bigl( \Delta c^{\dagger}_{i\uparrow}c^{\dagger}_{i\downarrow}
+\text{H.c.}\Bigr)
+J\sum_{i\sigma,\sigma^{\prime}}c^{\dagger}_{i\sigma}(\sigma_z)_{\sigma,\sigma^{\prime}}c_{i\sigma^{\prime}},\label{model}
\end{eqnarray}
with 
\begin{equation}
T=\left( \begin{array}{cc}
-t & \lambda_R/2 \\
-\lambda_R/2& -t 
\end{array} \right).
\end{equation}
Following similar ways in  Appendices A and B, we obtain following $4\times4$ BdG Hamiltonian after Fourier transformation with the basis $c_{k_x}=(c_{k_x\uparrow},c_{k_x\downarrow},c^{\dagger}_{-k_x\uparrow},c^{\dagger}_{-k_x\downarrow}):$
\begin{equation}
\mathcal{H}=\frac{1}{2}\sum_{k_x}c_{k_x}^{\dagger}\mathcal{H}(k_x)c_{k_x},
\end{equation}
\begin{equation}
\mathcal{H}(k_x)=\left( \begin{array}{cccc}
\xi_{k_x}+J&i\lambda_R\sin k_x&0&\Delta\\
-i\lambda_R\sin k_x&\xi_{k_x}-J&-\Delta&0\\
0&-\Delta&-\xi_{k_x}-J&-i\lambda_R\sin k_x\\
\Delta&0&i\lambda_R\sin k_x&-\xi_{k_x}+J
\end{array} \right).
\end{equation}
$\xi_{k_x}$ is the same as in Eq.(\ref{xi}).
The Matsubara Green's function is
\begin{equation}
G(\omega_n,k_x)=\frac{1}{i\omega_n-\mathcal{H}(k_x)}=\left( \begin{array}{cc}
G&F\\
\tilde{F}&\tilde{G}\\
\end{array} \right)\label{green02}.
\end{equation} 
By using Eq.(\ref{formula}) with 
\begin{eqnarray}
X=\left( \begin{array}{cc}
i\omega_n-\xi_{k_x}-J&-i\lambda_R\sin k_x\\
i\lambda_R\sin k_x&i\omega_n-\xi_{k_x}+J\\
\end{array} \right), \notag\\Y=\left( \begin{array}{cc}
0&-\Delta\\
\Delta&0\\
\end{array} \right),Z=\left( \begin{array}{cc}
0&\Delta\\
-\Delta&0\\
\end{array} \right),\notag\\
W=\left(\begin{array}{cc}
i\omega_n+\xi_{k_x}+J&i\lambda_R\sin k_x\\
-i\lambda_R\sin k_x&i\omega_n+\xi_{k_x}-J\\
\end{array} \right),
\end{eqnarray}
the anomalous part of the Matsubara Green's function, $F=\left(\begin{array}{cc}
f_{\uparrow\uparrow}&f_{\uparrow\downarrow}\\
f_{\downarrow\uparrow}&f_{\downarrow\downarrow}\\
\end{array} \right)$, is calculated to be
\begin{equation}
F=\frac{\Delta}{\zeta(\omega_n,k_x)}\left(\begin{array}{cc}
2i\lambda_R\sin k_x(J-\xi_{k_x})&-\gamma(\omega_n,k_x)+2i\omega_nJ\\
\gamma(\omega_n,k_x)+2i\omega_nJ&-2i\lambda_R\sin k_x(J+\xi_{k_x})\\
\end{array} \right),
\end{equation}
with
\begin{equation}
\gamma(\omega_n,k_x)=\Delta^2+\xi_{k_x}^2+\lambda_R^2\sin^2k_x+\omega_n^2-J^2\label{gamma},
\end{equation}
\begin{equation}
\zeta(\omega_n,k_x)=\gamma(\omega_n,k_x)^2+4\omega_n^2J^2-4\lambda_R^2\sin^2k_x(\xi_{k_x}^2-J^2).\label{zeta}
\end{equation}
The spin-singlet component of pair amplitude is given by
\begin{equation}
\frac{f_{\uparrow\downarrow}-f_{\downarrow\uparrow}}{2}=\frac{-\Delta\gamma(\omega_n,k_x)}{\zeta(\omega_n,k_x)}\label{sing},
\end{equation}
and spin-triplet ones are
\begin{eqnarray}
f_{\uparrow\uparrow}&=&\frac{2i\Delta\lambda_R\sin k_x(J-\xi_{k_x})}{\zeta(\omega_n,k_x)}\label{uu}\\
\frac{f_{\uparrow\downarrow}+f_{\downarrow\uparrow}}{2}&=&\frac{2i\omega_nJ}{\zeta(\omega_n,k_x)}\\
f_{\downarrow\downarrow}&=&\frac{-2i\Delta\lambda_R\sin k_x(J+\xi_{k_x})}{\zeta(\omega_n,k_x)}.\label{dd}
\end{eqnarray}
From Eqs.(\ref{xi}), (\ref{gamma}), (\ref{zeta}), and (\ref{sing}), the symmetry of spin-singlet pair amplitude is even-frequency and even-parity. Also, from Eqs.(\ref{xi}), (\ref{gamma}), (\ref{zeta}), and (\ref{uu})$\sim$(\ref{dd}), the symmetries of spin-triplet pair amplitudes, $f_{\uparrow\uparrow}$ and $f_{\downarrow\downarrow}$ are even-frequency and odd-parity, and that of the other component of spin-triplet, $\frac{f_{\uparrow\downarrow}+f_{\downarrow\uparrow}}{2}$, is odd-frequency and even-parity. We summarize these results in Table V. 
\begin{table}[htb]
\begin{tabular}{|c|c|c|c|c|c|c|c|c|c|c|c|} \hline
\multicolumn{3}{|c|}{\text{OTE} }& \multicolumn{3}{|c|}{\text{ESE} }& \multicolumn{3}{|c|}{\text{OSO} }& \multicolumn{3}{|c|}{\text{ETO} }\\ \hline\hline
$\uparrow\uparrow$&$\uparrow\downarrow$&$\downarrow\downarrow$& \multicolumn{3}{|c|}{$\uparrow\downarrow$}&\multicolumn{3}{|c|}{$\uparrow\downarrow$}&$\uparrow\uparrow$&$\uparrow\downarrow$&$\downarrow\downarrow$\\\hline
$\times$&$\bigcirc$&$\times$& \multicolumn{3}{|c|}{$\bigcirc$}&\multicolumn{3}{|c|}{$\times$}&$\bigcirc$&$\times$&$\bigcirc$\\\hline
\end{tabular}
\caption{Symmetries of the pair amplitude of the infinite ferromagnetic chain with Rashba-type SOC similar to table III.}
\end{table} 
\bibliography{nanowire1}
\end{document}